\documentclass[12pt]{article}

\usepackage{scicite}
\usepackage{times}

\usepackage{color,gensymb,amsmath,amssymb,graphicx,morefloats}
\usepackage{bm}
\usepackage{graphicx,morefloats}
\usepackage[resetlabels]{multibib}
\usepackage{subfigure} 
\usepackage{color,soul}
\usepackage{caption}

\newcites{SM}{Supplementary References}

\definecolor{burgundy}{rgb}{0.5, 0.0, 0.13}
\definecolor{burgundy2}{rgb}{0.5, 0.0, 0.5}

\newcommand{\be}[0]{\begin{equation}}
\newcommand{\ee}[0]{\end{equation}}
\newcommand{\beq}{\begin{eqnarray}}
\newcommand{\eeq}{\end{eqnarray}}

\newcommand{\dn}[0]{\downarrow}
\newcommand{\up}[0]{\uparrow}

\begin{document}

\thispagestyle{empty}

\hyphenation{va-ni-sh-ing}



\baselineskip24pt

\begin{center}
\vspace{-1.5cm}

{\Large 
Topological semimetals without quasiparticles
}
\\[0.6cm]

\normalsize{Haoyu Hu$^{1}$, Lei Chen$^{1}$, Chandan Setty$^{1}$, 
Mikel\ Garcia-Diez$^{2,3}$, Sarah\ E.\ Grefe$^{4}$, \\
Andrey\  Prokofiev$^5$, 
Stefan Kirchner$^{6,7}$, 
Maia\ G.\ Vergniory$^{2,8}$,
Silke Paschen$^{5,1}$, \\
Jennifer Cano$^{9,10}$, and Qimiao Si$^{1,\ast}$}
\\
[0.1cm]

\small\it{$^1$Department of Physics and Astronomy, Rice Center for Quantum Materials, Rice University,
Houston, Texas, 77005, USA
\\[0.1cm]

$^2$Donostia International  Physics  Center,  P. Manuel  de Lardizabal 4,  20018 Donostia-San Sebastian,  Spain
\\[0.1cm]

$^3$Department of Physics, University of the Basque Country (UPV-EHU), Bilbao, Spain
\\[0.1cm]

$^4$Theoretical Division, Los Alamos National Laboratory, Los Alamos, New Mexico 87545, USA\\[0.1cm]

$^5$Institute of Solid State Physics, TU Wien, Wiedner Hauptstr.\ 8-10, 1040 Vienna, Austria\\[0.1cm]

$^6$Department of Electrophysics, National Yang Ming Chiao Tung University, Hsinchu 30010, Taiwan
\\[0.1cm]

$^7$Center for Emergent Functional Matter Science, National Yang Ming Chiao Tung University, Hsinchu 30010, Taiwan
\\[0.1cm]

$^8$Max Planck Institute for Chemical Physics of Solids, Noethnitzer Str. 40, 01187 Dresden, Germany
\\[0.1cm]

$^9$Department of Physics and Astronomy, Stony Brook University, Stony Brook, NY 11794, USA\\[0.1cm]

$^{10}$Center for Computational Quantum Physics, Flatiron Institute, New York, NY 10010, USA\\[0.1cm]
}




\end{center}


The interplay between interactions and topology in quantum materials is of extensive current interest. 
Strong correlations are known to be important for insulating topological states, 
as exemplified by the fractional quantum Hall effect.
For the metallic case, whether and how 
they can drive topological states that
have no free-electron counterparts
is an open and pressing question. 
We introduce a general framework for lattice symmetries to constrain single-particle
excitations even when they 
are not quasiparticles,
and substantiate it in a periodic Anderson model with two channels of conduction electrons. 
We demonstrate that
symmetry constrains correlation-induced emergent excitations to produce 
non-Fermi
liquid topological phases. The loss of quasiparticles in
these phases
is manifested in a non-Fermi liquid form of spectral and transport properties, 
whereas its topological nature is characterized by surface states and valley and spin Hall 
conductivities. We also identify candidate materials to realize the proposed phases.
Our work opens a door  to a variety of 
non-Fermi liquid topological phases 
in a broad range of strongly correlated
materials.

\newpage

States of matter and their transitions have traditionally been classified 
in terms of Landau's order-parameter framework,
which is based on global symmetry and its spontaneous breaking \cite{LANDAU1936}.
For free-electron systems, typically with large spin-orbit couplings, the topological paradigm 
distinguishes electronic states in terms of certain topological invariants associated with their quantum-mechanical wavefunctions \cite{berry_phase}.
The electronic topology is responsible for various novel phenomena, such as unusual Hall effects 
and development of nontrivial surface states. In recent years, much progress has been made
in realizing metallic topological phases, including those that harbor Weyl and Dirac 
fermions \cite{Armitage2017,Nagaosa2020}. Crystalline symmetries play an important role in stabilizing
such topological semimetals
\cite{Bradlyn2017,Cano2018,Po2017,Watanabe2017,bradlyn2016beyond,cano2021band}.

In metallic topological states, electron correlations are typically treated perturbatively
\cite{Armitage2017,Nagaosa2020}.
The pursuit of strongly correlated topological semimetals is largely in its infancy
\cite{Wit14.1,Pas21.1}.
Nonetheless, recent  studies -- both theoretical \cite{Lai2018,Grefe2020_2} and 
experimental \cite{Dzsaber2017,Dzsaber2021} -- have demonstrated that correlations
drive composite fermions with highly renormalized topological nodal excitations that are pinned to the 
Fermi energy. An important ingredient
in realizing the correlated topological semimetals is that strong correlations
and space-group symmetry 
cooperate
\cite{Chen2021topological}.
As exemplified by the fractional quantum Hall effect \cite{Sto99.1}, strong correlations can lead to 
insulating topological states that  do not have 
noninteracting counterparts. In the metallic case, 
realizing
such topological states 
represents an outstanding challenge.

\begin{figure}[ht!]


    \centering
    
    \includegraphics[width=1\textwidth]{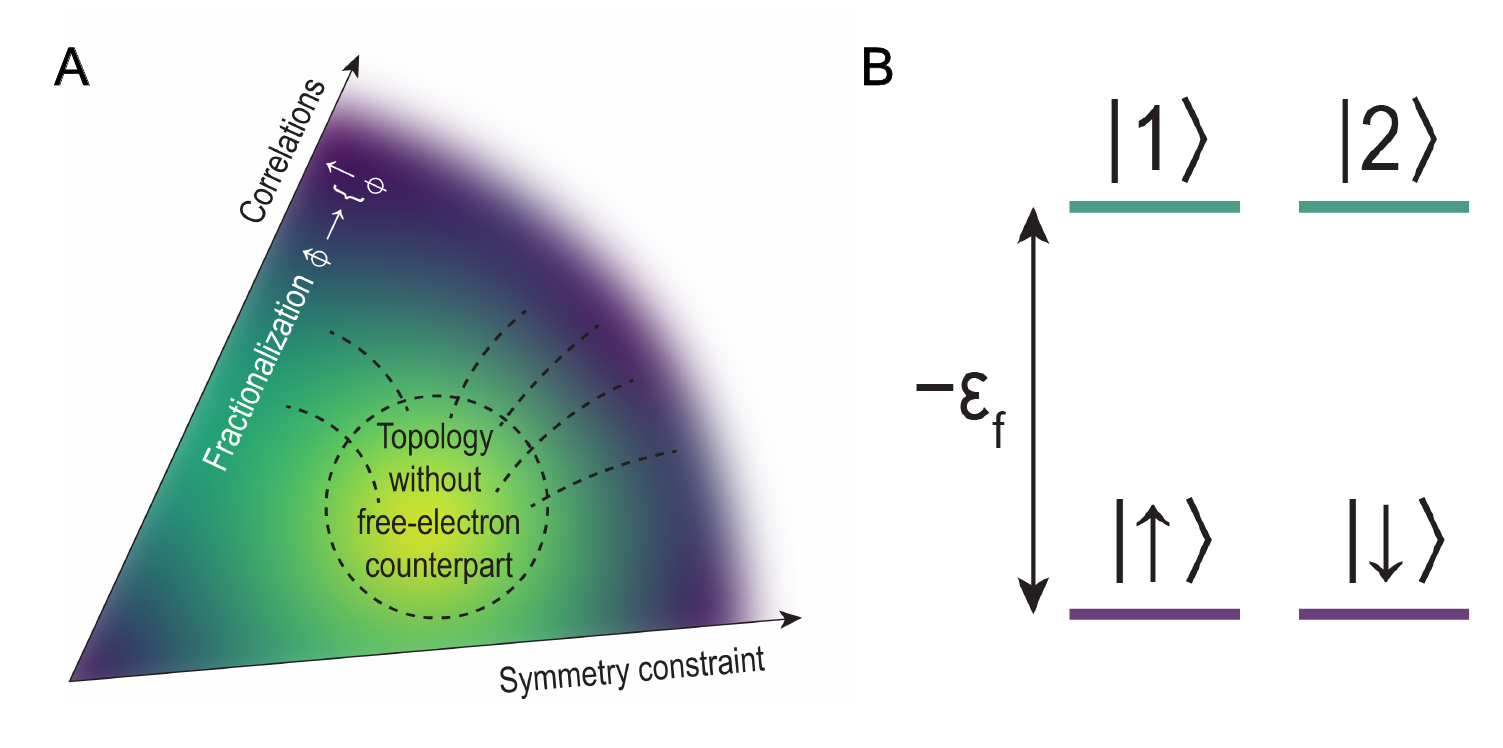}   
    \vspace{1cm}
 
   
    \caption{
    {\bf Illustration of the proposed route
    for
    electronic topological phases
    without quasiparticles.
    }
      ({\bf A}) An approach for topological phases without 
      free-electron counterparts:  Lattice symmetries constrain emergent excitations
      that are induced by strong correlations and particularly fractionalization.
    ({\bf B}) Strongly correlated local degrees of freedom involved in the two-channel Anderson lattice model.}
    \label{fig:1}

\end{figure}

Here we advance the notion that 
space-group symmetry constrains emergent  (low-energy)  electronic excitations, which are induced by
strong correlations and
particularly electron fractionalization, to produce non-Fermi liquid topological phases.
Such phases lose Landau quasiparticles and have no free-electron counterpart. 
Figure\,\ref{fig:1}A illustrates the idea.

We start by providing a general theoretical foundation for the lattice symmetries 
of interacting systems to constrain
dispersive electronic excitations, even when they are no longer quasiparticles.
The Green's functions describe the electronic (single-particle) excitations  of an interacting electron system.
The Green's function matrix $[G]$ can be expressed in terms of their 
eigenvectors, $v_{i}(\omega,\bm{k})$, and eigenfunctions
\cite{abrikosov2012};
the imaginary part of the eigenfunctions corresponds to the 
spectral functions, with their peaks describing dispersive electronic excitations
regardless of whether they are quasiparticles.
Consider the action of an element, $h$, of the 
space group $H$ on an electron annihilation operator $\psi$ in the space 
of internal quantum numbers ($a$),
frequency ($\omega$) and wave vector ($\bm{k}$):
\beq
h\in H:\,\,\psi_{a,i\omega,\bm{k}} \rightarrow \sum_b [U_h^\dag(\bm{k}) ]_{ab}\psi_{b,i\omega,h\bm{k}}\, ,
\eeq
where $U_h(\bm{k})$ is a unitary matrix. It follows from symmetry that 
$[\mathcal{U}_{h}v_{i}]_{(a,i\omega,\bm{k})}$ is also an eigenvector 
of 
$[G]$, where $
[\mathcal{U}_h]_{(a,i\omega,\bm{k}),(b,i\omega',\bm{k}')} = \delta_{\omega,\omega'}\delta_{\bm{k}',h\bm{k}}U_h(\bm{k})$. [The details are described in the Supplementary Materials (SM), Sec.\,A.]

We therefore reach the important conclusion that 
 the set of all eigenvectors $\{v_{i}\}_{i=1,2,...}$ form a representation 
space of the space group $H$.  This sets the stage to examine the dispersive modes through 
the subset of eigenvectors that form certain irreducible representations of the space group 
associated with a given wavevector 
$\bm{k}$, similarly as for the Bloch functions in the case of noninteracting electrons
\cite{cano2021band} 
(SM, Sec.\,A). 
As we tune the wavevector to some high symmetry points,
where additional symmetry leads to a two or higher dimensional irreducible representation,
the entire (frequency-dependent) spectral functions of the corresponding dispersive modes 
must be degenerate. The same conclusion is reached if, alternatively,
as we continuously tune the wavevector,
the peaks of two dispersive modes from two different irreducible representations cross;
this node cannot be removed by symmetry preserving
disruptions.
As such, our procedure provides a robust route to correlated topological semimetals, 
including ones that cannot be adiabatically connected to free-electron systems.

We now turn to a proof-of-principle demonstration of this general framework
in a well-defined model.
Our focus is on a periodic Anderson model, in which correlated
electrons act as local moments that are coupled to two degenerate channels of conduction electrons. 
As illustrated in  Fig.\,\ref{fig:1}B,
the local degrees of freedom at any given site correspond to a doublet 
in the ground state manifold, with (pseudo-)spins $\sigma=\up,\dn$, 
which hybridizes with two channels of conduction electrons, labeled by $\alpha=1,2$,
via an excited doublet.
Two channels of Kondo couplings lead to frustration in the development of 
Kondo entanglement, providing a means of fractionalizing the electronic excitations
\cite{stefan2020,Aff93.1,Eme92.1,Mal97.1}.
The model, thus, provides an ideal setting to explore lattice symmetry constraints on emergent
excitations  for non-Fermi liquid topology.
Further details of the model and solution method are provided in the SM
(Secs.\,B,C,D).

To address the symmetry constraints on emergent excitations, we choose to
analyze the model in a three-dimensional (3D)
kagome lattice, where the space group allows for protected nodes and, 
in addition, whose inversion symmetry breaking can be readily tuned (SM, Secs.\,B,C).
Here, $c_{i,\sigma,\alpha}$ ($d_{i,\sigma,\alpha})$ denotes
a conduction (local) electron,
describing the physical $spd$- ($f$-) electron,
on site $i$,
which marks both the unit cell position and sub-lattice index. 
We solve  the model within dynamical mean-field theory (DMFT) \cite{kot04},
while ensuring that our results capture the {\it asymptotically-exact}
 $\sqrt{|\omega|}$ low-energy behavior
in the multi-channel Kondo model~\cite{Cox93,Aff93.1}. 
The Green's function of the conduction electrons are calculated via  
$G_{c}(\omega,\bm{k})=[\omega-t_{\bm{k}} - \Sigma_c(\omega)]^{-1}$.
Here, $t_{\bm{k}}$ is the hopping matrix,
and $\Sigma_c(\omega)$ is the local self-energy that 
encodes  the correlation effects 
on the single-electron excitations.
We solve the dynamical equations in real frequency and reach 
extremely low temperatures 
(compared to $T_K$, the Kondo temperature, which is the 
dynamically generated low-energy scale)
(SM, Sec.\,D)
in order to reveal the asymptotic low-energy behavior.

\begin{figure}[ht!]
    \centering
    
    \includegraphics[width=1\textwidth]{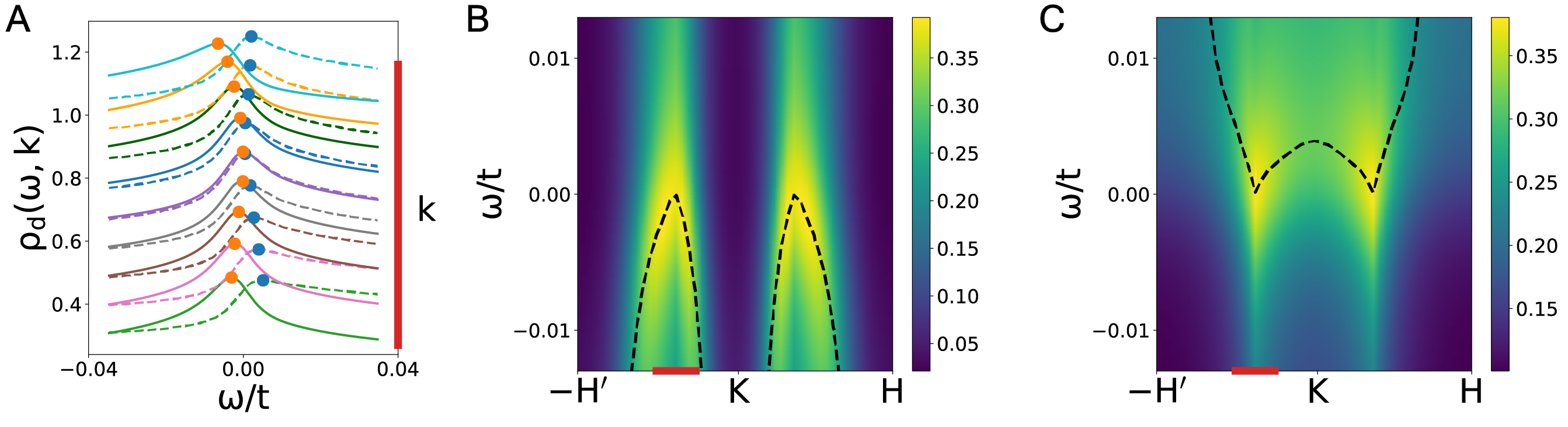}  
    \vspace{1cm}
 
    
    \caption{ {\bf Emergence of Weyl nodal excitations 
    beyond quasiparticles.}
    ({\bf A}) The $d$ electron spectral functions 
    ($\rho_d$)
    at various $\bm{k}$
     points 
    (marked by the red bar on the right) along a high symmetry line in the 3D kagome lattice. 
    Solid and dashed lines denote two dispersive modes from two eigenvalues of the Green's function.
    The blue and orange dots label the positions of the spectral peaks, which meet at a Weyl point.
    Each $\rho_d$
    curve
     has been shifted vertically to avoid overlapping. 
    ({\bf B}),({\bf C}) illustrate the spectral functions of the two Weyl-point-bearing branches. Here,
     the dashed line denotes the energy spectral peaks.
     The red bar marks the cut of wave vectors in the Brillouin zone 
     along which the spectral functions 
     are shown in panel {\bf A}.
    The calculations are done at the temperature
     $T=10^{-3}T_K$.}
    \label{fig:3}
\end{figure}

 The imaginary part of the conduction-electron self-energy, ${\rm Im} \Sigma_c(\omega)$,
 is found to depend on frequency in a square-root fashion in the low-energy limit
 (SM, Fig.\,S2A).
 The $\sqrt{|\omega|}$ form ensures that the 
 spectral width of any dispersive mode will become progressively
 larger than its energy on approach of the Fermi energy,
 in contrast to what happens when the electronic excitations are quasiparticles.
It describes
 a non-Fermi liquid,
 as characterized by an electrical resistivity that depends on temperature ($T$) in a
 $\sqrt{T}$ (SM, Fig.\,S2B).
 These asymptotic results are found only when our calculations are done at 
 sufficiently low temperatures 
 (SM, Sec.\,E).
The asymptotic form of the self-energy, however, is not adequate:
On its own, it would not lead to new dispersive modes 
near the Fermi energy.
 Importantly, we find that 
 the self-energy 
 is described by a pole
 in the complex energy plane, 
 with relevant energy scales of the order $T_K$.
 This pole,
 in turn,
 gives rise to narrow bands that are pinned to the Fermi energy 
 (Secs.\,E,F and Fig.\,S6).
The results 
for the narrow bands are shown in Fig.\,\ref{fig:3}A. 
 
In the above sense, the dispersive modes near the Fermi energy are {\it emergent} electronic 
excitations. Crucially, they go beyond Landau quasiparticles:
For reasons described earlier and as seen in  Fig.\,\ref{fig:3}A,
 any peak of the spectral function  {\it vs.} energy
 for a given wavevector $\bm{k}$
  has a width that is always
  {\it larger} than the peak energy as measured 
 from the Fermi energy; 
 this persists on approach of the Fermi energy,
 and is opposite to what happens with Landau quasiparticles.
The disperse modes capture the development 
in the spectral weight of the physical $f$-electrons (represented in the model by the $d$-operators)
 in the immediate vicinity of the Fermi energy
(SM, Fig.\,S4A): they appear within a range of the Kondo energy.

What is remarkable is that these 
dispersive modes form Weyl nodes
which are bound near to the Fermi energy, 
 as seen in 
 Fig.\,\ref{fig:3}A 
 (along the high symmetry line, $-H'$-$K$-$H$, of the Brillouin zone,
 as marked by the red line that is specified in Figs.\,\ref{fig:3}B,C).
 They 
 comprise two close-by branches, which are illustrated more clearly in 
 Fig.\,\ref{fig:3}B and 
 Fig.\,\ref{fig:3}C, respectively. To understand the Weyl nodes, 
 we analyze the structure of the interacting Green's function in terms of the Green's function 
 eigenvectors and eigenfunctions. 
 As  discussed earlier 
 (and in the SM, Sec.\,A),
 the eigenvectors are labeled by the wave vector 
 $\bm{k}$
 and
  are subject to the same crystalline symmetry constraints as Bloch functions. 
  Indeed,
 the two dispersive modes shown in Fig.\,\ref{fig:3}A have 
 different symmetry eigenvalues and form a symmetry-protected node
 (see the SM, Sec.\,H);
 a similar mechanism produces additional non-Fermi-liquid Dirac nodes 
along the high symmetry line $-A$-$\Gamma$-$A$.
In other words, 
in this model,
a robust mechanism underlies the emergence of the Weyl nodes and the associated
nodal excitations
that go beyond quasiparticles.
 
\begin{figure}[ht!]
 
    \centering
    \includegraphics[width=1\textwidth]{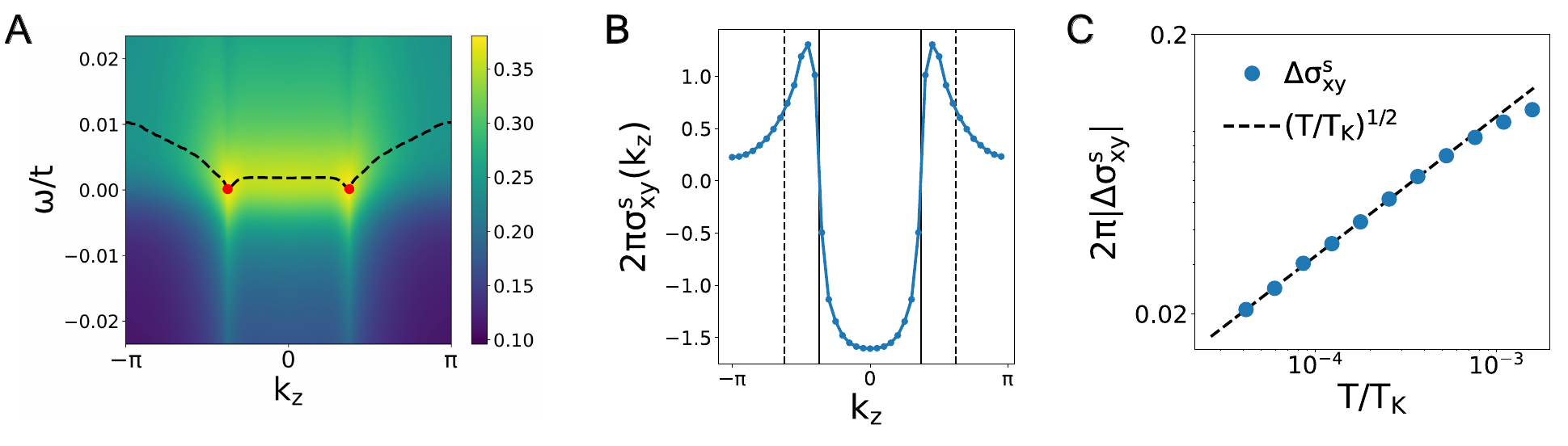}    
     \vspace{1cm}

       
    \caption{
    {\bf Characterizing the topological nature of the correlated phase.}
    ({\bf A}) The spectrum of the edge mode at $k_x=4 \pi/3$ in the 3D kagome lattice at $T/T_K=10^{-3}$ with 
    an open boundary along the $y$ direction;
    the two red dots mark
    the Weyl nodes; 
    the dashed line denotes the energy spectral peak.
    ({\bf B}) In-plane spin-Hall conductivity $\sigma_{xy}^s$ 
    as a function of $k_z$ in the 3D kagome model at $T/T_K=10^{-3}$. 
    The vertical solid/dashed lines mark
     the $k_z$ coordinates of the planes in which the 
     Weyl/Dirac nodes reside. 
     A rapid change (in the form of a smeared jump) occurs
     when $k_z$ crosses the 
     Weyl/Dirac-node-hosting planes.
    ({\bf C})
    The total spin Hall conductivity of the 3D kagome model showing a $\sqrt{T}$ dependence. Here, the extrapolated zero-temperature value ($\sim 0.25$) is subtracted in order to highlight the nature of the $T$-dependent component.
    }
    \label{fig:4}
\end{figure}

We characterize the topological nature of these non-Fermi liquid 
phases in several ways. We start by analyzing the surface states. Their
spectral functions are plotted in Fig.\,\ref{fig:4}A; their exponential decay 
from the surface to the interior of the system is demonstrated in the SM (Sec.\,I and
Fig.\,S8A).
In addition, 
in Fig.\,\ref{fig:4}B,
we 
show the in-plane spin-Hall conductivity as a function of $k_z$
(see the SM, Sec.\,J),
which implicates the topological nature of the Weyl nodes. 
The overall spin-Hall conductivity has a $\sqrt{T}$ dependence, 
as shown in 
Fig.\,\ref{fig:4}C.  
At the same time, the nonlinear Hall effect is found to display a singular $\frac{1}{\sqrt{T}}$ 
dependence
below the 
 Kondo temperature scale
(see the SM, Sec.\,K).
We also expect the valley Hall conductivity
to have a $\sqrt{T}$ dependence,
as explicitly calculated for a model 
that shows
Dirac nodal states
without quasiparticles
(see the SM, Fig.\,S7E). 
The surface states and the various Hall effects 
also
provide the means
 to identify 
the topological semimetal phases
experimentally.

Having advanced 
genuinely interacting Weyl semimetal phases,
we now turn to their 
materials
realization.
 So far we have analyzed the two-channel 
 periodic
 Anderson model on the 3D kagome lattice,
by assuming the existence of the pertinent local degrees of freedom as
 illustrated in 
 Fig.\,\ref{fig:1}B. From the model studies, we gain the key insight
that the form of the electron self-energy
 is insensitive to the underlying lattice, 
provided the required local degrees of freedom 
are in place; for example, 
its form on the 2D kagome or 3D cubic lattice is essentially the same (see 
the SM, 
Fig.\,S5) as what we have described for the 3D kagome lattice case. 
We thus look for space group symmetries 
that constrain nodes in the dispersion while also allowing for the local degrees of freedom of 
Fig.\,\ref{fig:1}B.

This prescription has led us to three materials, the centrosymmetric 
PrBi \cite{PrBi_exp}
 and
PrFe$_4$P$_{12}$ \cite{PrFe4P12_exp}
as well as the noncentrosymmetric 
UPt$_3$Au$_2$ \cite{Qui88.1}, with the space groups
$\#225$,
$\#204$, and $\#216$, 
respectively, and the point group symmetries $O_h$, $T_h$, and $T_d$ at the Pr/U site. 
The Pr$^{3+}$ and U$^{4+}$ ions have the electronic configuration $4f^2$ which, for these point group symmetries, 
{\em can} lead
to a $\Gamma_3$ non-Kramers doublet ground state. 
A non-Kramers doublet can be treated as pseudo-spin degree of freedom. The first
excited levels are then Kramers doublets, whose spin degrees of freedom provide
the pseudo-channel  index \cite{Cox98}. PrBi has a low charge carrier
concentration \cite{PrBi_exp}, PrFe$_4$P$_{12}$ can be readily tuned to such
a state \cite{PrFe4P12_exp}, and UPt$_3$Au$_2$ \cite{Qui88.1} exhibits
 a very weakly temperature-dependent resistivity, all suggestive of semimetallic ground states  
(see 
the SM, Sec.\,L, for further details). 
Our
analysis of space group symmetry constraints suggests the emergence of symmetry
protected Dirac nodes in PrBi and PrFe$_4$P$_{12}$ and symmetry protected Weyl nodes in UPt$_3$Au$_2$
(see
the SM, Sec.\,M).
We have verified
our symmetry argument
in the representative case of PrBi:
Its electronic structure of the non-$f$ conduction electrons (SM,
Sec.\,N) is shown
in Fig.\,S11;
symmetry-protected Dirac points away from the Fermi energy are indeed found 
 along the $\Gamma$-$X$ line
 (which also implies that
 the Kondo-generated emergent excitations 
in PrBi
 will have Dirac nodes near the Fermi energy).
Thus, to realize the proposed correlated gapless topological phases, we put forward the three
as candidate materials for systematic low-temperature studies in their respective phase diagrams.

We close with 
several important remarks.
Firstly, we have already discussed the robustness of our theoretical results for the 
two-channel Kondo/Anderson lattice models, in that we recover the asymptotically exact low-energy
behavior of the self-energy while also 
uncovering the emergence of a pole in the 
self-energy. Furthermore, the low-temperature non-Fermi liquid behavior we have 
presented in the
models is consistent with experimental results
observed in some (generic-lattice, in particular cubic) heavy-fermion metals 
that host non-Kramers doublets in the ground-state manifold and, hence, two-channel Kondo lattices
\cite{Sakai2011}. 
Secondly, the loss of quasiparticles in our case is associated with emergent 
nodal states created by the Kondo effect 
near the Fermi energy. This is
in contrast to the perturbative 
effect of interactions on 
pre-existing nodal states of conduction electrons \cite{Armitage2017,Nagaosa2020,Moon13}.
Thirdly, there can be additional broken-symmetry states at low temperatures, 
following the
paradigm in strongly correlated systems
that non-Fermi liquids nucleate novel 
phases \cite{Pas21.1}.
For generic lattices, the multi-channel 
Kondo
 correlations may lead to broken-symmetry phases such as superconductivity \cite{Tsujimoto2014},
multipolar order \cite{Sakai2011} or even
time-reversal-symmetry-broken
``hastatic" order \cite{Chandra2013}.
 It will be instructive to explore
any new physics that the nontrivial topology brings in these states.
Fourthly,
the role of electron 
fractionalization 
can be further explored in the proposed non-Fermi liquid topological semimetals.
The two-channel Kondo effect
fractionalizes electrons into 
Majorana fermions \cite{Mal97.1},
which underlie the loss of Landau quasiparticles. 
Finally, we have emphasized the constraints of lattice symmetries 
on dispersive modes associated with peaks of the electron spectral functions. 
Because 
it
 keeps track of both energy and wave vector, our approach 
will likely allow
for constraining the nature of topological states with other 
types of features in correlations,
such as zeros of Green's functions in topological Mott states
\cite{Morimoto2016,Mai22.1}.

To summarize, our work has advanced the first 
topological semimetal
that has emergent electronic topological excitations beyond
quasiparticles and that has
no free-electron counterpart. 
We have advanced a general 
framework,
{\it viz.} to place symmetry constraints on the dispersive modes of single-particle Green's functions,
which allows us to establish the development of non-Fermi liquid topological semimetals in a robust way.
In specific lattices,
we have shown how their space group symmetries 
constrain
the 
two-channel Kondo correlations to produce Weyl (and Dirac)
nodal states that are pinned near the Fermi energy 
but are not quasiparticles. 
Guided by the considerations of both symmetry and strong correlations,
we have proposed several materials to realize these
non-Fermi liquid topological semimetals. 
The proposed approach has the potential to help realize
gapless topological phases without 
 free-electron counterparts in a variety of other strongly correlated materials and structures.


\bibliographystyle{Science}
\bibliography{Weyl2Ch_ref}

\noindent{\bf Acknowledgments:}
Work at Rice has been supported by the 
the National Science Foundation under Grant No.\ DMR-2220603 (L.C. and Q.S.),
the Air Force Office of Scientific Research under Grant No.\ FA9550-21-1-0356 (C.S.)
and the Robert A. Welch Foundation Grant No.\ C-1411 (H.H.).
The majority of the computational calculations have been 
performed on the Shared University Grid at Rice funded by NSF under Grant EIA-0216467, 
a partnership between Rice University, Sun Microsystems, and Sigma Solutions, 
Inc., the Big-Data Private-Cloud Research Cyberinfrastructure MRI-award funded by NSF under Grant No.\ CNS-1338099,
and the Extreme Science and Engineering Discovery Environment (XSEDE) by NSF under Grant No.\  DMR170109.
M.G.-D. and M.G.V. acknowledge support from the Spanish Ministry of Science and Innovation 
Grant No. PID2019-109905GB-C21 and the Deutsche Forschungsgemeinschaft (DFG, German Research Foundation) GA 3314/1-1 -- FOR 5249 (QUAST).
Work at Los Alamos was carried out under the auspices of the U.S. Department of Energy (DOE) National 
Nuclear Security Administration under Contract No.\ 89233218CNA000001, 
and was supported by LANL LDRD Program.
A.P. and S.P. acknowledge funding from the Austrian Science Fund (projects No.\ I4047 and 29279), the European  Microkelvin  Platform  (H2020 project No.\ 824109)
and the FWF via the Research unit QUAST-FOR5249.
S.K. was in part supported 
by the Ministry of Science and Technology, Taiwan (grant No. MOST 111-2634-F-A49-007) 
and the Center for Emergent Functional Matter Science of National Yang Ming Chiao Tung University from ``The Featured Areas Research Center Program" within the framework of the 
``Higher Education Sprout Project by the Ministry of Education (MOE)" in Taiwan, by
the Yusha Fellowship Program of the  MOE  Taiwan, 
and by the National Key R\&D Program of the MOST of China, 
Grant No.\ 2016YFA0300202 and the National Science Foundation of China, Grant No.\ 11774307.
J.C. acknowledges the support of
 the National Science Foundation under Grant No.\ DMR-1942447 
and the support of the Flatiron Institute, a division of the Simons Foundation.
Four of us (S.G., S.P., J.C. and Q.S.) acknowledge the hospitality of the Aspen Center for Physics,
which is supported by NSF grant No. PHY-1607611.

\noindent{\bf Authors contributions:}
All authors contributed to the research of the work and the writing of the paper.

\noindent{\bf Competing interests:}
The authors have no competing interests.

\noindent{\bf Data availability:}
The data that support the findings of this study are available from the corresponding author
upon reasonable request.

\noindent$^{\ast}$To whom correspondence should be addressed; E-mail:  qmsi@rice.edu.

\newpage


\clearpage

\newpage


\clearpage

\newpage

\clearpage

\newpage

\setcounter{figure}{0}
\setcounter{equation}{0}
\makeatletter
\renewcommand{\thefigure}{S\@arabic\c@figure}
\renewcommand{\theequation}{S\arabic{equation}}

\noindent{\LARGE\bf{Supplementary Materials}}
\\

\noindent{\bf\Large
Topological
semimetals without quasiparticles
}

\noindent Haoyu\ Hu, Lei\ Chen, Chandan\ Setty, Mikel\ Garcia-Diez,
Sarah\ E.\ Grefe, 
Andrey\  Prokofiev,\\
Stefan\ Kirchner, Maia\ G.\ Vergniory, Silke\ Paschen,
Jennifer\ Cano, and Qimiao\ Si$^{\ast}$\\

\noindent$^{\ast}$To whom correspondence should be addressed; E-mail:  qmsi@rice.edu.\\

\noindent Materials and Methods\\
Figs. S1--S11\\
References (36--60, see above)
 
\section*{Materials and Methods}

\subsection*{A. Symmetry of Green's function: Space group symmetry constraints on interacting systems }
 We study the symmetry properties of the 
  Green's functions:
   \begin{eqnarray*}
 [G(\tau,\bm{r},\tau',\bm{r'})]_{ab} = -\langle T_\tau \psi_{a,\bm{r},\tau}  \psi^\dag_{b,\bm{r},\tau'}\rangle \, ,
 \end{eqnarray*}
where $\psi_{a,\bm{r},\tau}^\dag$ creates an electron with index $a$, which characterizes its spin, 
orbital and sublattice, at position $\bm{r}$ and imaginary time $\tau$.
We will first prove that, in a quantum phase with symmetry of the space group $H$, 
the eigenvectors of the Green's function form a representation of both $H$ 
and the corresponding little group $H_{\bm{k}}$. Based on these symmetry properties, 
we can analyze the symmetry-protected crossing of single-particle excitations.

In general, it is more convenient to work with Matsubara frequency and wave vector by introducing Fourier transformation of electron operators:
\begin{eqnarray*}  
\psi_{a,i\omega,\bm{k}} &=& \frac{1}{\beta N}\sum_{\bm{r}}\int_0^{\beta} \psi_{a,\bm{r},\tau} e^{i\omega\tau -i\bm{k}\cdot \bm{r} }d\tau \, ,
\end{eqnarray*}
where $N$ is the total number of unit cells and $\beta$ is the inverse temperature. The Fourier transformation leads to following Green's function, $[G]_{(a,i\omega,\bm{k}), (b,i\omega',\bm{k'})} = -\langle \psi_{a,i\omega,\bm{k}}\psi_{b,i\omega',\bm{k'}}^\dag \rangle $,
which can be treated as a matrix with row index $(a,i\omega,\bm{k})$ and column index $(b,i\omega',\bm{k'})$. As a matrix, we can introduce the eigenvectors and eigenvalues:
\begin{eqnarray*}
\frac{1}{N\beta}\sum_{b,i\omega',\bm{k'}}[G]_{(a,i\omega,\bm{k}), (b,i\omega',\bm{k'})}[v_{i}]_{(b,i\omega',\bm{k'})} = g_{i}[v_{i}]_{(a,i\omega,\bm{k})}\, , 
\end{eqnarray*}
where $g_{i}$ and $[v_{i}]_{(a,i\omega,\bm{k})}$ denote the $i$-th eigenvalue and $i$-th eigenvector.

We then prove the set of all eigenvectors $\{v_{i}\}_{i=1,2,...}$,
form a representation space of the space group $H$ whose element is defined as
\begin{eqnarray*}
h\in H&:&\psi_{a,i\omega,\bm{k}} \rightarrow \sum_b [U_h^\dag(\bm{k}) ]_{ab}\psi_{b,i\omega,h\bm{k}}\, ,
\end{eqnarray*}
where $U_h(\bm{k})$ is a unitary matrix. The symmetry of the system requires the Green's function to satisfy
\begin{eqnarray*}
[\mathcal{U}_h^{\dag}\,G\,\mathcal{U}_h]_{ (a,i\omega,\bm{k}), (b,i\omega',\bm{k'}) } =[G]_{ (a,i\omega,\bm{k}), (b,i\omega',\bm{k'}) } \, ,   
\end{eqnarray*}
where 
$
[\mathcal{U}_h]_{(a,i\omega,\bm{k}),(b,i\omega',\bm{k}')} = \delta_{\omega,\omega'}\delta_{\bm{k}',h\bm{k}}U_h(\bm{k})
$. Consequently, $[\mathcal{U}_{h}v_{i}]_{(a,i\omega,\bm{k})}$ is also an eigenvector 
of the Green's function matrix $[G]$ and we can then introduce the matrix representation 
$\mathcal{D}[h]$ of $h\in H$ via
\begin{eqnarray*}
[\mathcal{U}_h \, v_{i}]_{a,i\omega,\bm{k}} = \sum_j \mathcal{D}[h]_{ij} \, [v_{j}]_{a,i\omega,\bm{k}}\,.
\end{eqnarray*}
Therefore, the set of all eigenvectors $\{v_{i}\}_{i=1,2,...}$ form a representation of space group $H$ with matrix representation $\mathcal{D}[h]$.

In practice, the 
systems we deal with hold
 space and time translation symmetry, which allows us to decompose $G$ into block diagonal forms 
\begin{eqnarray*}
[G]_{(a,i\omega,\bm{k}), (b,i\omega',\bm{k'})} = \delta_{\omega,\omega'} \delta_{\bm{k},\bm{k'}} [G(i\omega,\bm{k})]_{ab}\, .
\end{eqnarray*}
We also perform an analytical continuation by replacing $i\omega$ with $\omega+i0^+$ and work with the retarded Green's function. 
Due to the block diagonalized structure, it's enough to consider the eigenvalues and eigenvectors of each block characterized by $(\omega,\bm{k})$~\cite{sym_green, sym_green_2,abrikosov2012}:
\begin{eqnarray*}
\sum_b [G(\omega,\bm{k})]_{ab}[v_{i}(\omega,\bm{k})]_b = g_{i}(\omega,\bm{k})[v_{i}(\omega,\bm{k})]_a \, ,
\end{eqnarray*}
where $g_{i}(\omega,\bm{k})$ and $[v_{i}(\omega,\bm{k})]_b$ are the $i$-th eigenvalue and $i$-th eigenvector of each block. We define the $i$-th mode in terms of the combination of $g_{i}(\omega,\bm{k})$ and $v_{i}(\omega,\bm{k})$. The corresponding spectral function of the $i$-th mode is 
\beq 
\rho_{i}(\omega,\bm{k}) =-\frac{1}{\pi}\text{Im}[ g_{i}(\omega,\bm{k})] \, .
\nonumber 
\eeq 
The full spectral function is then the sum of spectral functions of each mode
\beq 
\rho(\omega,\bm{k}) = -\frac{1}{\pi}\text{Tr}\text{Im}[G(\omega,\bm{k})] = \sum_i \rho_{i}(\omega,\bm{k}).
\nonumber 
\eeq 

To analyze the symmetry properties of the dispersive modes, it's useful to consider the little group $H_{\bm{k}} \leq H$. For a given $\bm{k}$ in the first Brillouin zone, the little group $H_{\bm{k}}$ is defined as $H_{\bm{k}} = \{ h \in H| h\bm{k} = \bm{k}\}$~\cite{cano2021band}. Since the symmetry operation $h\in H_{\bm{k}}$ won't change the wave vector $\bm{k}$, we can focus on the set of eigenvectors $\{v_{i}(\omega,\bm{k})\}_{i=1,2,...}$ with the wave vector index $\bm{k}$. 
As we mentioned earlier, the eigenvectors form a representation space of the space group $H$. Thus, the eigenvectors of the $\bm{k}$ block, $\{v_{i}(\omega,\bm{k})\}_{i=1,2,...}$, form a representation space 
of the little group $H_{\bm{k}}$, which is also a subgroup of $H$. 
The matrix representation $D_{\omega,\bm{k}}[h]$ for $h\in H_{\bm{k}}$ is defined via
\beq 
 [U_h(\bm{k}) v_{i}(\omega,\bm{k})]_a = \sum_{j}\bigg[D_{\omega,\bm{k}}[h]\bigg]_{ij} [v_{j}(\omega,\bm{k} )]_a\, .
\eeq 
We can further decompose the representation into a direct sum of irreducible representations and analyze their symmetry properties.
In this way, the lattice symmetry operates in interacting systems on the eigenfunctions of the Green's function in parallel to how it acts in noninteracting systems on the Bloch functions
\cite{Bradlyn2017,Cano2018,Po2017,Watanabe2017,bradlyn2016beyond,cano2021band}.

Based on the representation decomposition, we can inspect the dispersive modes of each irreducible representation. As we continuously tune $\bm{k}$, if the peaks of two dispersive modes from two different irreducible representations cross, then such crossing cannot be gapped out by symmetry preserved perturbations. Another type of symmetry-protected crossing could appear when we tune the $\bm{k}$ 
to certain high symmetry points, where additional symmetry leads to a two or higher dimensional irreducible representation. In this case, the spectral functions of the corresponding dispersive modes 
must degenerate. 
A detailed explanation of the symmetry-protected Weyl/Dirac node in our microscopic model is 
provided in this SM (Sec.\,H). 

\subsection*{B. Periodic Anderson model with two channels of conduction electrons}

We consider a two-channel periodic Anderson model.
While such a model can be considered 
in a generic setting 
for correlation physics,
we focus on two 
representative
lattices
in which crystalline symmetries promote Dirac and Weyl nodes. 
At every site, the local degrees of freedom (Fig.\,\ref{fig:1}B)
hybridize with two channels 
of conduction electrons, labeled by $\alpha=1,2$;
the hybridization involves the change of local electron number by $1$, which 
turns
the local
configuration to $|\alpha = 1,2 \rangle$, and likewise from $|\alpha\rangle$ to $|\sigma\rangle$.
In practice, one way of realizing the two-channel periodic Kondo/Anderson model is 
to use an orbital degree of freedom (non-Kramers doublet such as $\Gamma_3$) to 
represent the pseudo-spin $\sigma$ leaving the physical spins of the conduction electrons and 
a higher Kramers doublet to represent the channel $\alpha$ \cite{Cox98,Sakai2011}.
The single-impurity version of this model, the two-channel Anderson/Kondo
model, is well established to have non-Fermi liquid behavior
 \cite{stefan2020,nozieres1980kondo,Cox98,Aff93.1,Pustilnik04,Keller2015,Iftikhar2015,Yeh2020};
 the electron correlations fractionalize the $d$-electron, leading to
 an emergent Majorana fermion \cite{Eme92.1,Mal97.1}.
 
  We consider both 3D and 2D kagome lattices, where the 3D kagome lattice corresponds to the stacking of the kagome planes with two added ingredients -- the inversion symmetry is broken and
a non-zero spin orbit coupling reduces the $SU(2)$ spin symmetry down to the $U(1)$ spin symmetry
(see Sec.\,C). The model can be expressed in terms of the following Hamiltonian:
 \beq \label{eq:lattice_ham}
{\cal H}&=&\sum_{ij,\sigma,\alpha}(t_{ij,\sigma\sigma'}-\mu\delta_{ij}\delta_{\sigma\sigma'})c_{i,\sigma,\alpha}^\dag c_{j,\sigma',\alpha} \nonumber  \\
&&+ 
P\bigg[ \epsilon_f \sum_{i,\sigma,\alpha}d_{i,\sigma,\alpha}^\dag d_{i,\sigma,\alpha} 
+ V\sum_{i,\sigma,\alpha}( c_{i,\sigma,\alpha}^\dag d_{i,\sigma,\alpha} +\text{h.c.}) \bigg]P 
\, .
\eeq
Here,
$c_{i,\sigma,\alpha}$ ($d_{i,\sigma,\alpha})$ 
denotes
a conduction (local) electron 
of spin $\sigma$ and channel $\alpha$ on site $i$, 
which marks both the unit cell position and sub-lattice index.
The projection operator $P$
allows for
the strongly correlated $d$-electrons to have
two empty states $|1\rangle,|2\rangle$ and two singly occupied states $|\up\rangle,|\dn\rangle$
at every site. In addition, $t_{ij,\sigma\sigma'}$ is the hopping matrix of the conduction electrons, with $\mu$ acting as the chemical potential;
$\epsilon_f$ 
specifies the spacing of the $d$-electron levels (Fig.\,\ref{fig:1}B),
and
$V$ is the hybridization strength between the two types of electrons. 
We solve the 
model
within dynamical mean-field theory (DMFT) \cite{kot04} 
as described in this SM, Sec.\,D;
importantly, we utilize
a dynamical large-$N$ method that captures 
the asymptotic low-energy behavior of the two-channel Kondo model \cite{Cox93,Cai20_charge_Kondo}
(see Sec.\,D).

\subsection*{C. Noninteracting band structure} 

The noninteracting Hamiltonian of the 3D kagome lattice is 
\beq
{\cal H}_0 &= &\sum
\begin{bmatrix}
c_{\bm{k},\alpha,\up}^\dag &c_{\bm{k},\alpha,\dn}^\dag 
\end{bmatrix}
\begin{bmatrix} 
h(k) -\mu \mathbb{I}& \\ 
&& h^*(-k)-\mu \mathbb{I}
\end{bmatrix} 
\begin{bmatrix}
c_{\bm{k},\alpha,\up} \\ c_{\bm{k},\alpha,\dn} 
\end{bmatrix} \nonumber \\
c_{\bm{k},\alpha,\sigma} &= &\begin{bmatrix}
c_{A,\bm{k},\alpha,\sigma} & c_{B,\bm{k},\alpha,\sigma} & c_{C,\bm{k},\alpha,\sigma} 
\end{bmatrix} ^T   \, .
\eeq 
Here, $A,B,C$ denote three sublattices located at
 $(\frac{1}{4},\frac{\sqrt{3}}{4},0)a,(\frac{1}{2},0,0)a,(\frac{1}{4},-\frac{\sqrt{3}}{4},0)a$ respectively
(thus, each unit cell has three sites).
The hopping matrix in the spin up sector is:
\begin{eqnarray*}
h(k)
&=&[t-t_{z,2}\cos(k_z)]\\
&&\begin{bmatrix}
0&[(1+\alpha) + (1-\alpha)e^{-ik_2} ]  
& [(1+\alpha)e^{ik_1} +(1-\alpha)e^{-ik_2}]  \\
[(1+\alpha) + (1-\alpha)e^{ik_2} ] & 0& [(1+\alpha) e^{ik_1}+(1-\alpha) ] \\
[(1+\alpha)e^{-ik_1} +(1-\alpha)e^{ik_2}]  & [(1+\alpha) e^{-ik_1}+(1-\alpha) ] & 0
\end{bmatrix} \\
&&-t_z \cos(k_z)\begin{bmatrix}
1 &0 & 0 \\
0 & 1 & 0 \\
0 & 0 & 1 
\end{bmatrix}-t_2
\begin{bmatrix}
0&[e^{i(-k_1-k_2)}+e^{ik_1}] 
& (1+e^{i(k_1-k_2)}) \\
[e^{i(k_1+k_2)}+e^{-ik_1}] & 0& [e^{i(k_1+k_2)}+e^{-ik_2}]\\
[1+e^{-i(k_1-k_2)}] &  [e^{-i(k_1+k_2)}+e^{ik_2}] & 0
\end{bmatrix}   \\ 
&&+[i\lambda + i\lambda_z \cos(k_z)] 
\begin{bmatrix}
0&  [1+e^{-ik_2}] &[-e^{ik_1}-e^{-ik_2}] \\
- [1+e^{ik_2}]& 0& [1+e^{ik_1} ]\\
 [e^{-ik_1}+e^{ik_2}]  &- [1+e^{-ik_1} ] &0 
\end{bmatrix}  
\, , 
\end{eqnarray*}
where
\begin{eqnarray*}
k_{1/2}&=&\frac{1}{2}(k_x \pm \frac{k_y}{\sqrt{3}} ) \, .
~~~~~~~~~~~~~~~~~~~~~~~~~~~~~~~~~~~ 
~~~~~~~~~~~~~~~~~~~~~~~~~~~~~~~~~~~ 
~~~~~~~~~~~~~~~~~~~~~~~~~~~~~~~~~~~ \, .
\end{eqnarray*}
In addition, $t,t_2$ are the nearest-neighbor and next-nearest-neighbor intra-layer hopping parameters, 
and $t_z$ and $t_{z,2}$ 
are the nearest-neighbor and next-nearest-neighbor inter-layer hopping elements. 
$\alpha$ denotes the anisotropy of $t_1$ and $t_{z,2}$ hopping that breaks mirror symmetry $M_y$ and inversion symmetry. 
$\lambda$ and $\lambda_z$ 
are the intra-layer and inter-layer spin orbit couplings, respectively, which reduce the $SU(2)$ spin symmetry down to $U(1)$.
The corresponding 2D kagome model without spin-orbit coupling and inversion symmetry breaking can be realized 
by setting $t_z=t_{z,2}=\lambda=\lambda_z=0$, $\alpha=0$ and dropping all the terms that contain $k_z$. 

In practice, we set $t_1=0.29,t_2=0.14,t_z=-0.17,t_{z,2} = 0.26,\lambda=0.43,\lambda_z=0.06,\alpha=1.4,\mu=-1.09$ for the 3D kagome lattice. 
As shown in Fig.\,S1B, Fig.\,S1E and Fig.\,S1F, we observe two Weyl nodes at $(1/3,1/3,z)$, $(2/3,2/3,-z)$, 
two anti-Weyl nodes at $(1/3,1/3,-z)$, $(2/3,2/3,z)$, and two Dirac nodes at $(0,0,\pm z')$ where $z\approx 0.19,z'\approx 0.32$. 
Weyl and Dirac nodes are all protected by $C_{3z}$ symmetry. 
In Fig.\,S1E and Fig.\,S1F, we label the relevant bands by their eigenvalues of $C_{3z}$ symmetry, 
where $\eta=e^{i2\pi/6}$.
The $C_{3z}$ symmetry excludes the hybridization between the two bands with different eigenvalues, 
and thus protects the nodes.
The bands along the high symmetry line ($-A$-$\Gamma$-$A$) are two-fold degenerate due to the mirror symmetry $M_z$ and time-reversal symmetry.
 The eigenvalues inside (outside) the brackets are for the spin-up (spin-down) sector.
Because the hybridization between spin-up and spin-down electrons is forbidden by the $U(1)$ spin symmetry, there can be no gap opened by hybridizing two bands with the 
same $C_{3z}$ eigenvalue but opposite spin indices.

For the 2D kagome lattice, we set $t_1=1, t_2=0.3,\mu=-1.0$. 
(We have chosen generic parameters so that the band structure is relatively simple.)
The noninteracting dispersion and Dirac node are shown in Fig.\,S1D. 
At the high symmetry point $K/K'$, three bands form $\Gamma_1 \oplus \Gamma_3$ representations of the little group $C_{3v}$, 
where $\Gamma_3$ is a 
two-dimensional 
irreducible representation and thus protects the Dirac node. Here, due to the $SU(2)$ symmetry, we consider the spinless representations of the group.

\subsection*{D. Solution methods:
Dynamical mean-field theory and the dynamical large-$N$ 
approach
}

The two-channel Kondo effect is a robust mechanism for non-Fermi liquid physics \cite{Aff93.1}.
Here, we describe the dynamical method that captures the asymptotic low-energy form 
of the non-Fermi liquid  behavior \cite{Cox93,Cai20_charge_Kondo}, 
while incorporating the lattice symmetries that can constrain the emergent electronic excitations.

To solve the lattice model, 
we first introduce the auxiliary-particle representation of the projected local electron annihilation 
operator: 
$Pd_{i,\sigma,\alpha}P = f_{i,\sigma}b_{i,\alpha}^\dag$. Here, $f_{i,\sigma}$ 
is a pseudo-fermion annihilation operator 
and $b_{i,\alpha}^\dag$ is a 
pseudo-boson creation operator. The Hilbert space of the 
$d$ electron can be expressed as: $|\sigma\rangle = f_\sigma^\dag |\text{vac}\rangle$, 
$|\alpha\rangle =b_\alpha^\dag |\text{vac}\rangle$ 
with the vacuum state $|\text{vac}\rangle$. There is also a local constraint 
on the pseudo-particle occupation numbers,
 $Q_i = 
 \sum_{\sigma}f_{i,\sigma}^\dag f_{i,\sigma} + \sum_{\alpha} b_{i,\alpha}^\dag b_{i,\alpha} 
=1 $.

In DMFT, the local correlators of the two-channel periodic Anderson model is determined
by the self-consistent local model \cite{Georges_2chAL1992}:
\beq \label{eq:impurity_action}
S_{loc}&=& \int \sum_{\sigma,\alpha}c_{\sigma,\alpha}^\dag (\tau)  G_0(\tau,\tau')^{-1} c_{\sigma,\alpha}(\tau') d\tau d\tau' \nonumber \\
&+&\int \Bigg[
\sum_\sigma f_\sigma^\dag(\tau)(\partial_\tau - \epsilon_f) f_\sigma(\tau) +\sum_\alpha b_\alpha^\dag(\tau)\partial_\tau b(\tau) \bigg] d\tau \nonumber \\
&+& \int V\sum_{\sigma,\alpha} (c_{\alpha,\sigma}^\dag(\tau) f_{\sigma}(\tau)b_{\alpha}^\dag(\tau) +h.c.) d\tau \nonumber\\
&+&\int i\lambda(\tau) \bigg(\sum_\sigma f_\sigma^\dag(\tau) f_\sigma(\tau) +\sum_\alpha b_\alpha^\dag(\tau) b_\alpha(\tau) -Q\bigg) d\tau  
\, .
\eeq  
Here, $\lambda$ is a Lagrangian multiplier enforcing the pseudo-particle constraint. Whereas
 $G_0$ is the bath function of the conduction electrons, which
  is determined from the DMFT equations: 
\beq 
&&G_{c,loc}(\omega) = \frac{1}{L^d}\sum_{\bm{k}} \frac{1}{\omega - t_{\bm{k}} -\Sigma_c(\omega)} \nonumber \\
&&G_{0}^{-1}(\omega) - G_{c,loc}^{-1}(\omega) = \Sigma_c(\omega) \, .
\eeq 
$G_{c,loc}(\omega)$ is the local Green's function of 
the $c$ electrons, 
$\Sigma_c(\omega)$ is the associated self-energy which is taken to be local, i.e., 
$\bm{k}$-independent within the DMFT. 
We note that the effective
local model in the 3D lattice model 
has an effective $SU(2)$ spin symmetry. The bath functions for the 
spin up and spin down conduction electrons are the same due to the time reversal symmetry and no hybridization between spin up and spin down electrons is allowed given the $U(1)$ spin symmetry. 

The local model is solved via 
the saddle point equations \cite{Cox93,Kroha1999FermiAN,NCA,2CK_nrg}
of a dynamical large-$N$ limit
 [with a generalization to an $SU(N)$ symmetry for the local electrons' spins and an 
 $SU(N)\times SU(M)$ symmetry for the conduction electrons, while
 taking the large-$N,M$ limit with a fixed $N/M=1$]. 
Here, the self-energies of the pseudo-fermion and pseudo-boson are 
\beq 
&&\Sigma_f(\omega) = MV^2 \int \rho_0(\omega-\epsilon) f(\epsilon-\omega) G_b(\epsilon) d\epsilon  \nonumber \\ 
&&\Sigma_b(\omega) = NV^2 \int \rho_0(\epsilon-\omega) f(\epsilon-\omega) G_f(\epsilon) d\epsilon \nonumber 
\eeq 
where $\rho_0$ is the density of states associated with the bath function $G_0$ and $f(x)$ is the
 Fermi-Dirac function. In conjunction with the Dyson equations for the pseudo-particle propagators $G_f$,$G_b$,
\beq 
&&G_f^{-1}(\omega)=\omega-\epsilon_f-\lambda -\Sigma_f(\omega) \nonumber\\
&&G_b^{-1}(\omega) = \omega -\lambda - \Sigma_b(\omega)  \nonumber\, ,
\eeq 
we have a complete set of equations. The local Green's function is calculated via $G_{c,loc}(\omega)=G_0(\omega) +V^2G_0(\omega)G_d(\omega)G_0(\omega)$, with $G_d(\omega)$ 
being the correlator of the $d$ electron.
We note that $V$ is at the order of $1/\sqrt{N}$ in the large-$N$ approach.
 These equations are known to capture the asymptotic 
low-energy behavior -- including the non-Fermi liquid exponents -
of the multi-channel Kondo/Anderson models \cite{Cox93}. Importantly, the $1/N$ vertex corrections do not change 
the asymptotic low-energy behavior \cite{Cox93,Cai20_charge_Kondo}.

In practice, we take $V=1.0,\epsilon_f=-0.5$ for the 2D kagome model and $V=1.5,\epsilon_f=-1.0$ for 
the 3D kagome model. 
We work on the retarded Green's functions and solve the DMFT 
self-consistent equation by varying the bath function $G_{c,0}$ and stop 
the procedure
when the difference between two iterations is less then $1\%$. 

We solve the dynamical large-$N$ equations in real frequency. It turns out to be important to
reach sufficiently low temperatures 
(down to about $10^{-3}-10^{-5}$ of the Kondo temperature $T_K$)
in order to reveal the asymptotic low-energy behavior.
We find $T_K/t \approx 0.03$ for the model on the 
3D kagome lattice and $T_K/t\approx 0.01$ for the model on the 2D kagome lattice.

In 
Fig.\,S3A and Fig.\,S3B, 
we show the local spectral functions of  the 
pseudo-bosons and pseudo-fermions of the 3D kagome lattice. We observe an $\omega/T$ scaling 
with
\begin{eqnarray}
\rho_f(\omega,T) &= (\frac{T}{T_K})^{1/2}g_f(\omega/T) \nonumber \\
\rho_b(\omega,T) &= (\frac{T}{T_K})^{1/2}g_b(\omega/T) \, ,
\end{eqnarray} 
where $g_f(x),g_b(x)$ are universal functions. 
The results for the 2D kagome lattice (not shown) are very similar.
 In addition, 
to illustrate the power-law 
dependence, Fig.\,S3C and Fig.\,S3D depict
 $\rho_b$ vs. frequency at a low temperature for the 3D and 2D kagome lattices, respectively.

\subsection*{E. Breakdown of quasiparticles}
 The imaginary part of the conduction-electron self-energy, ${\rm Im} \Sigma_c(\omega)$,
 is found to depend on frequency in a square-root fashion in the low-energy limit, 
 as shown in
 Fig.\,S2A. 
 This is in contrast to the Fermi liquid case, where ${\rm Im} \Sigma_c(\omega)$ 
 goes quadratically in $\omega$. The square-root frequency dependence of the self-energy describes
 a non-Fermi liquid, with a loss of quasiparticles. The non-Fermi liquid behavior is
 characterized by an electrical resistivity that depends on temperature ($T$) in a
 $\sqrt{T}$  way,  which is illustrated in 
 Fig.\,S2B. 
In practice, we find that the asymptotic behavior, in both the frequency and temperature dependences,
 appears only at sufficiently low temperatures; it was not found in previous DMFT solutions 
 of the multichannel Kondo lattice model on the infinite dimensional hypercubic lattice \cite{Jar96}
solved at much higher temperatures (about $0.1$ $T_K$ and above).

 The self-energy can be described by a form that contains both the 
 asymptotic low-energy behavior and a pole
 in the complex energy
 plane.
 \beq 
\Sigma_c(\omega) \sim A \frac{\gamma}{\omega-\omega_0 + i\gamma} + g(\omega) \, \, \Theta(\Lambda-|\omega|) \, ,
\label{eq:sigc_MainText}
\eeq
where
$\Theta(x)$ is the Heaviside step function, and 
$g(\omega) \sim a + c\sqrt{|\omega|}$ describes 
the low-energy asymptotic behavior that appears below the energy cutoff $\Lambda$. 
(For further details, see Sec.\,F.)
To our knowledge, the existence of a pole in the self-energy has 
not been recognized 
in any two-channel Kondo model; its uncovering is made possible by our working 
on the real-frequency axis and solving the
model at very low temperatures. 
Physically, the pole captures the development of the $f$-electron spectral 
weight within an energy range of $T_K$
near the Fermi energy.
Because the $f$-electron state can decay into the fractionalized (Majorana) excitations, 
the damping rate 
$\gamma$ 
remains nonzero even at the zero-temperature limit.

 The structure of the self-energy implies
 two branches of
 dispersive modes near the Fermi energy $E_F$,
 with velocities highly reduced by the factor $\sim T_K/D$, where $D$ is the bare bandwidth
(see Sec.\, F). 
 
 \subsection*{F. Self-energy: asymptotic low energy behavior, a pole and the resulting emergent  dispersive modes}

In this section, we describe the nature of the self-energy 
$\Sigma_c$ in some detail.
Due to the symmetry requirement,
 the local self-energy $\Sigma_c$ is diagonal in spin, sublattice and channel spaces, which simplifies the analysis.
As shown in Fig.\,S4A,
$\text{Im}[\Sigma_c](\omega)$ has a peak near the Fermi energy. 
We find that this peak is 
well
represented by a Lorentzian function;
we define the  peak center and width as $\omega_0$ and $\gamma$, respectively.
Accordingly, we
obtain $\Sigma_c(\omega)\sim A \frac{\gamma}{\omega-\omega_0+i\gamma}$.
In addition, we observe a $\sqrt{|\omega|}$ behavior in the low energy limit as shown 
in Fig.\,S2A. Such a dependence
fully captures the asymptotic low-energy behavior
for the two-channel Kondo model~\cite{Cox93,Aff93.1}.
Combining both contributions, we have the
 form
of the self-energy
given in the 
previous section
(Eq.\,\ref{eq:sigc_MainText}).

As shown in Fig.\ S4A, we fit the imaginary part of the self-energy with this form.
The Heaviside step function has been replaced by a soft cutoff via a hyperbolic tangent function
in the fitting.
In 
Fig.\ S4B, we also plot the evolution of $\gamma$, which characterize the width of the peak, 
as a function of temperature. At low temperature, we find that $\gamma$ saturates to a finite value at a fraction of 
 $T_K$.
 This saturation further supports
 the robustness of the pole in the complex plane.

Physically, the Kondo effect always leads to the development of 
$f$-electron spectral weight within an energy range of $T_K$
near the Fermi energy. In the single-channel Kondo/Anderson lattice,
such spectral weight is carried by Landau quasiparticles;
a pole in $\Sigma_c(\omega)$ develops whose decay rate must 
satisfy the Fermi liquid form and depends on $\omega$ in an $\omega^2$ form
(where $\omega$ is measured from the Fermi energy).
By contrast, in the two-channel Kondo/Anderson lattice model,
the $f$-electron state can decay into the fractionalized (Majorana) excitations.
Consequently, the damping rate 
$\gamma$ 
remains nonzero even at the zero-temperature and zero-frequency limit.

The existence of a pole in
 the  self-energy leads
 to a band reconstruction in the interacting model, as shown in Fig.\,S6. 
Combining the self-energy 
and symmetry group, we are able to analyze the structure of the Green's function near the 
Dirac and Weyl nodes. 

To further explicate the consequence of the pole in the self-energy, we
 stress that
 emergent excitations develop 
 at low energies (within the Kondo scale near the Fermi energy) out of this pole.
They take the form of hybridized bands, except that these are not quasiparticles. Specifically,
in the vicinity of
 the crossing wave vector $\tilde{k}$, we have 
\begin{equation}
\tilde{P}G^{-1}_c(\omega,\tilde{k}+k)\tilde{P} \approx 
U^\dag \begin{bmatrix}
\omega-E_0 -v_zk_z-\Sigma_c(\omega) & -v_xk_x+iv_yk_y \\
-v_xk_x-iv_yk_y  & \omega-E_0+v_zk_z-\Sigma_c(\omega)   \\
\end{bmatrix} 
U\nonumber 
\end{equation} 
 where $\tilde{P}$ is the projection matrix that only keeps the relevant bands,
  $U$ is a unitary matrix that brings the dispersion near a Weyl/Dirac node
   into the canonical form and $E_0$ is the energy of the noninteracting Weyl/Dirac nodes. 
   
 Importantly, a pole of $\Sigma_c(\omega)$ in the complex-frequency plane leads to a {\it doubling} of 
 the bands, creating dispersive modes within the Kondo energy $T_K$ near the Fermi energy that are not quasiparticles.
  We illustrate this explicitly for the case of
 Dirac nodes in 2D, where we can set $v_z=0$
 and 
 keep implicit the two-fold degeneracy in spin space. 
 The corresponding spectral functions would have {\it low-energy} peaks  at 
 \beq
 \omega \approx \omega_0+ {\gamma} \frac{ A}{E_0}(1\pm \sqrt{v_x^2k_x^2+v_y^2k_y^2+v_z^2k_z^2}/E_0)
 \eeq
 We re-iterate that this equation describes the {\it emergent} dispersive modes located
 near Fermi energy. The results are showin in Fig.\,7A,B,C. The corresponding results
 for the 3D kagome lattice have already been shown in the main text, Fig.\,2A,B,C.
 
 Recognizing that both $E_0$ and ${ A}$ are non-universal and generically of the order of the 
 bare conduction electron bandwidth $D$, and $\gamma$ is of the order of $T_K$,
the velocity  of the emergent Weyl nodal states is renormalized down from its noninteracting counterpart
by a factor of $\sim (T_K/D)$.
Each peak corresponds to one branch of the Dirac/Weyl node in the
non-Fermi liquid dispersive modes.

In the broad energy range,  
the  pole in the self-energy 
 essentially leads 
to a band reconstruction in the interacting model, as shown in Fig.\,S6A,B.
The corresponding results for the 3D kagome lattice are shown in Fig.\,S6C,D.

\subsection*{G. Topological Dirac semimetal without quasiparticles in the 2D model
}
The Dirac spectrum appearing near the Fermi energy, for the case of the 2D kagome lattice, 
is shown in Fig.\,S7\,A,B,C.
The emergent dispersive modes near the Fermi energy form
a symmetry-protected Dirac node at the $K$ point. 
The non-Fermi liquid properties 
in the 2D kagome lattice are also characterized by the $\sqrt{T}$ dependence in both 
the resistivity and valley Hall conductivity, as shown in Fig.\,S7\,D,E. 

\subsection*{H. Symmetry of Green's function in the microscopic models}
Based on the eigenvectors and eigenvalues of the single-particle Green's function, we now prove that the Weyl (Dirac) nodes in the 3D kagome model and 
the Dirac nodes in the 2D kagome model are both protected by crystalline symmetry, 
but via different mechanisms. Since single-particle excitations are composite fermions, it is adequate
 to focus on the Green's function of the $d$ electrons, $G_d.$

The Weyl nodes in 3D kagome model come from the band inversion mechanism. Along the high symmetry line $K-H$, we can assign an eigenvalue of $C_{3z}$ to each dispersive mode (eigenvectors of Green's function).
 In Fig.~\ref{fig:3}A, two modes with left-moving 
 and right-moving 
 peaks 
 have symmetry eigenvalues $\eta^3$ and $\eta^5$, respectively. 
The different eigenvalues would protect the crossing of the two modes
under symmetry-preserved perturbations. 
A similar mechanism protects the Weyl nodes on $-H$-$K'$-$H'$ and Dirac nodes on $-A$-$\Gamma$-$A$.

 In the 2D kagome lattice, the Dirac nodes are symmetry enforced. $\{v_{d,i}(\omega,\bm{k}=K/K')\}_i$ 
 forms the $\Gamma_1\oplus \Gamma_3$ representation of the little group $G_{K/K'} = C_{3v}$. Due to the $SU(2)$ symmetry, 
 we can ignore the spin index and use a spinless representation here. The two-dimensional irreducible representation $\Gamma_3$ 
 enforces a two-fold-degenerate (four-fold-degenerate 
 including the spin degree of freedom) spectrum  at the $K/K'$ point.
 Consequently, the two dispersive modes have to coincide with each other at the $K/K'$ point as shown in 
 Fig.\,S7A.

\subsection*{I. Edge modes}
To evaluate the edge mode in the 3D kagome model, we take a slab geometry with 50 unit cells
 and open boundary along $y$ direction.
In the interacting model, we consider the Green's function $[G_d(\omega,k_x,k_z)]_{y_1\alpha_1,y_2\alpha_2}$ that is 
a $300\times 300$ matrix, where $y_i$ is the coordination of unit-cell 
along the $y$ direction and $\alpha_i$ is the sublattice and spin index. 
Through diagonalizing the Green's function matrix, 
we obtain the eigenvalues $\{g_{i}(\omega,k_x,k_z)\}_i$ and 
eigenvectors $\{[v_i(\omega,k_x,k_z)]_{y\alpha}\}_i$. For a local self-energy, 
$[v_i(\omega,k_x,k_z)]_{y\alpha}$ 
becomes $\omega$ independent and describes the real-space distribution of the corresponding mode. Numerically, 
we identify an edge mode that decays exponentially in the real space as shown in Fig.\,S8A. The corresponding spectral function $\text{Im}[g_i(\omega,k_x,k_z)]/\pi$ is shown in the main text,
Fig.\,3A. 
    
\subsection*{J. Hall conductivities}
In this section, we describe the calculation of the spin and valley Hall conductivities.
Within the DMFT approach, in which vertex corrections are formally suppressed, the in-plane spin Hall conductivity of each $k_z$: $\sigma^s(k_z)$, can be expressed as 
\beq 
\sigma_{xy}^s(k_z) = \frac{1}{4\pi^2}\int_{k_x,k_y}\sum_{m,n\in \uparrow } \tilde{\Omega}_{mn}(k_x,k_y,k_z) 
- \frac{1}{4\pi^2}\int_{k_x,k_y}\sum_{m,n\in \downarrow } \tilde{\Omega}_{mn}(k_x,k_y,k_z)
\eeq 
where $m,n$ are band indices. The first (second) terms sum over bands with spin up (down).  
The full bulk contribution
is obtained via $\sigma_{xy}^s = \int_{k_z} \sigma_{xy}^s(k_z)$. In addition, 
\beq 
\tilde{\Omega}_{mn}(\bm{k}) &=&i\lim_{\nu\rightarrow 0}
\frac{d}{d(i\nu)}\frac{1}{\beta} \sum_{i\omega} \frac{(V_{\bm{k}}^\dag \partial_{k^x}t_{\bm{k}} V_{\bm{k}} )_{nm} (V_{\bm{k}}^\dag \partial_{k^y}t_{\bm{k}} V_{\bm{k}})_{mn}}{[i\omega+i\nu-\epsilon_{\bm{k},n}-\Sigma_{c}(i\omega)][i\omega -\epsilon_{\bm{k},m}-\Sigma_c(i\omega) ] } \nonumber \\
&=&i\lim_{\nu\rightarrow 0}
\frac{d}{d(i\nu)}\frac{1}{\beta}\sum_{i\omega}\frac{  [A^x_{\bm{k},mn}A^y_{\bm{k},nm}-A^x_{\bm{k},nm}A^y_{\bm{k},mn}](\epsilon_{\bm{k},n}-\epsilon_{\bm{k},m})^2 }{[i\omega+i\nu-\epsilon_{\bm{k},n}-\Sigma_{c}(i\omega)][i\omega -\epsilon_{\bm{k},m}-\Sigma_c(i\omega) ] } \,.
\nonumber
\eeq 
$V_k$, $\{\epsilon_{\bm{k},n}\}_{n}$ are the eigenvectors and eigenvalues
({\it i.e.} the Bloch bands)
 of the 
hopping matrix $t_{\bm{k}}$. $A^\mu_{mn,\bm{k}} = i\langle u_{m,\bm{k}}|\partial_{k^\mu} |u_{n,\bm{k}}\rangle$ is a Berry connection.

Similarly, we calculate valley Hall conductivity $\sigma^\nu$ of 2D kagome lattice via
\beq 
\sigma^{\nu} &= &\frac{1}{4\pi^2}\int_{|\bm{k}-K|<C }\sum_{m,n} \tilde{\Omega}_{mn}(\bm{k})-\frac{1}{4\pi^2} \int_{|\bm{k}-K'|<C } 
 \sum_{m,n}\tilde{\Omega}_{mn}(\bm{k}) ,
\eeq 
where the  $\bm{k}$ integration is 
 taken in a finite region near $K/K'$ with cutoff $C=\pi/2$.
 In the calculation, we introduce a small mass term to the Dirac point in order to generate a finite Berry curvature distribution. From top to bottom, we label the bands by 1, 2, 3.
 In the noninteracting limit, the lowest band that is below the Fermi energy contributes a positive $\sigma^\nu$. 
The top two bands are above the Fermi energy near $K(K')$ and, thus the corresponding contributions
$\tilde{\Omega}_{12}$ and $\tilde{\Omega}_{21}$ vanish.
Once we turn on the
hybridization
 $V$, 
dispersive modes
develop near the Fermi energy and generate nonzero $\tilde{\Omega}_{12}$ 
and $\tilde{\Omega}_{21}$. We can extract the contributions from $\tilde{\Omega}_{12}$ and $\tilde{\Omega}_{21}$ and define $\sigma^{\nu}_{12}$
\beq 
\sigma^{\nu}_{12} &= &\frac{1}{4\pi^2}\int_{|\bm{k}-K|<\Lambda } (\tilde{\Omega}_{12}(\bm{k})+\tilde{\Omega}_{21}(\bm{k}))-\frac{1}{4\pi^2} \int_{|\bm{k}-K'|<\Lambda } (\tilde{\Omega}_{12}(\bm{k})+\tilde{\Omega}_{21}(\bm{k}))\nonumber .
\eeq 
$\sigma^\nu_{12}$ is part of $\sigma^\nu$ that is directly generated by the interactions. The evolution of $\sigma^\nu$ and $\sigma^\nu_{12}$ 
as a function of hybridization 
is shown in Fig.\,S8B.

\subsection*{K. Non-linear Hall effect} 
We now turn to the non-linear Hall effect in the non-Fermi liquid phase. Using the Green's function 
approach~\cite{nqh}, the non-linear Hall conductivity is
\beq 
\chi_{xyy} &=& -\frac{1}{2\pi} \int_{\bm{k},\epsilon} \text{Im}\bigg\{\text{Tr}\bigg[\partial_\epsilon f(\epsilon)
u_x \partial_\epsilon G_c(\epsilon-i0^+)u_y G_c(\epsilon-i0^+)u_y G(\epsilon+i0^+) \nonumber 
\\
&&+\partial_\epsilon f(\epsilon)u_x\partial_\epsilon G_c(\epsilon -i0^+) u_{yy} G_c(\epsilon+i0^+) +f(\epsilon)
u_x \partial_\epsilon^2G_c(\epsilon-i0^+) u_{yy}\nonumber \\
&&+2f(\epsilon) u_x \partial_\epsilon[G_c(\epsilon-i0^+)u_yG_c(\epsilon-i0^+)]u_y G_c(\epsilon+i0^+) 
\bigg]
\bigg\} ,
\label{eq:non_linear_hall}
\eeq 
where $f(\epsilon)$ is the Fermi-Dirac function,  and
$u_{\mu_1\mu_2...\mu_n} = \partial^n_{k_{\mu_1}k_{\mu_2}...k_{\mu_n}}t_{\bm{k}}$. Each term contains a product of three Green's functions
 with two derivatives of $\epsilon$. In the asymptotic region $c_{IR}<\omega<<T_K$ where scaling behaviors appear, 
 $G_c(\epsilon) \sim g_0 + g_1\sqrt{|\epsilon|}$. $c_{IR}$ acts as an infrared cutoff. The integral behaves as $\sim \epsilon^{-3/2} d\epsilon$.
  (Note that at least one derivative is acting on the Green's function.)
   Transforming to the temperature behaviors, we would expect $\sim 1/\sqrt{T}$ 
    in the scaling region $c_{IR}<T<<T_K$.
To more explicitly observe this $\sim 1/\sqrt{T}$ behaviors, we consider the Green's function near the Weyl node $\tilde{\bm{k}}$
\beq 
G_c(\omega,\bm{k}+\tilde{\bm{k}})^{-1} \approx [\omega -h_0-\Sigma_c(\omega)]\mathbb{I} -h_x \sigma^x -h_y \sigma^y - h_z \sigma^z
\nonumber ,
\eeq 
where $\Sigma_c(\omega)$ is the self-energy. $\sigma^{x,y,z}$ are Pauli matrices in the $2\times 2$ subspace that holds the Weyl node. $h_0, h_x,h_y,h_z$ are functions of wave vector. At the Weyl node, $h_0(\tilde{\bm{k}})=E_0,h_x(\tilde{\bm{k}})=h_y(\tilde{\bm{k}})=h_z(\tilde{\bm{k}})=0$. The eigenvalues of 
the Green's function matrix are 
\beq 
&&g_{\pm}^{-1} = \omega - \epsilon_{\pm } -\Sigma_c(\omega) \nonumber \\
&&\epsilon_{\pm} = h_0 \pm \sqrt{h_x^2+h_y^2+h_z^2}
\nonumber ,
\eeq 
where $\Sigma_c(\omega)$ is the self-energy and has non-Fermi liquid behavior.
We also introduce the velocity matrix: $
v_\mu^{\alpha\gamma} = (V^\dag \partial_{k^\mu} t_{\bm{k}} V)_{\alpha\gamma} 
$, where $t_{\bm{k}}=h_0\mathbb{I}+h_x\sigma^x +h_y\sigma^y +h_z\sigma^z$ is the hopping matrix and $V$ is the matrix formed by the eigenvectors of the hopping matrix.

Here, we focus on the intrinsic contributions from the Berry curvature~\cite{nqh_nature_comm}. However, all the other contributions produce the same $1/\sqrt{T}$ behavior. The intrinsic contributions reads (from the first term of Eq.\,\ref{eq:non_linear_hall}), 
\beq
\chi^{in}_{xyy} = -\frac{1}{2\pi}\int_{\bm{k}} \int d\epsilon \frac{\partial f(\epsilon)}{\partial \epsilon} \text{Im}\bigg\{ \text{Tr} \bigg[&&
v_x^{+-}v_y^{-+}v_y^{++} \partial_\epsilon g_-(\epsilon+i0^+) g_+(\epsilon+i0^+)g_+(\epsilon -i0^+) \nonumber \\
&+&
v_x^{-+}v_y^{++}v_y^{+-} \partial_\epsilon g_+(\epsilon+i0^+) g_+(\epsilon+i0^+)g_-(\epsilon -i0^+)
\bigg] 
\bigg\} \nonumber .
\eeq  
Due to the pre-factor $\frac{\partial f(\epsilon)}{\partial \epsilon}$, it is sufficient to focus on the low-energy contribution. Since the self-energy behaves
 as $\text{Im}[\Sigma_c] \sim a + b \sqrt{|\omega|}$, approximately we have 
\begin{equation*}
\chi^{in}_{xyy} = -\frac{1}{2\pi} \int_k \int_\epsilon \frac{\partial f(\epsilon)}{\partial \epsilon} \Omega_{xy}^{+} v_y^{++} [a_0 + \frac{a_1}{ \sqrt{|\epsilon|}} +o(\epsilon)] 
\end{equation*}
where $a_1\approx \pi (\rho_0\frac{\sqrt{h_x^2+h_y^2+h_z^2}}{h_0})^3 \frac{1}{\sqrt{\Lambda} }$ and $\Lambda$ is the low energy scale of $\sqrt{|\omega|}$ behaviors and $\rho_0$ is the density of state near zero Fermi energy.
After integrating over $\epsilon$, we have 
\begin{equation}
\chi^{in}_{xyy} =  \frac{1}{\sqrt{T}}\cdot\bigg[\frac{-1.34}{2\pi}  \int_k \Omega_{xy}^+ v_y^{++}  a_1\bigg] +...
\end{equation}
where $\Omega_{xy}^+ = i\langle \partial_x u_+| \partial_y u_+\rangle - i\langle \partial_y u_+| \partial_x u_+\rangle $ is the Berry curvature.

\subsection*{L. Candidate materials}
We have identified three candidate materials, PrBi \cite{PrBi_exp,PrBi_dft,Wu19.3},
PrFe$_4$P$_{12}$ \cite{PrFe4P12_exp,PrFe4P12_theory}, and UPt$_3$Au$_2$ \cite{Qui88.1}, 
to
realize the 
correlated topological phases predicted here. They crystallize in
the space groups $\#225$, $\#204$, and $\#216$, respectively, which all allow
for a $\Gamma_3$ non-Kramers doublet ground state of the Pr$^{3+}$ or U$^{4+}$
$4f^2$ wave function. The former two are centrosymmetric and candidate
correlated Dirac semimetals, 
whereas the latter is noncentrosymmetric and thus a
candidate 
correlated Weyl semimetal.

In PrBi, evidence for a $\Gamma_3$ ground state was provided by fitting a
crystal electric field (CEF) model to temperature-dependent magnetic
susceptibility data; the assignment was supported by a much better agreement of
the specific heat data with this than other possible CEF level schemes
\cite{PrBi_exp}. A low charge carrier concentration was evidenced from quantum
oscillation experiments \cite{PrBi_exp}. For PrFe$_4$P$_{12}$, 
key evidence is summarized in Fig.\,S9. A non-magnetic CEF
ground state doublet is suggested by the antiferroquadrupolar ordering at low
temperatures (panel A) \cite{PrFe4P12_exp}. Strong Kondo coupling was evidenced
by a large Sommerfeld coefficient determined from specific heat data
\cite{Sug02.1}, and is confirmed by the large $A$ coefficient of the Fermi
liquid $T^2$ form seen in electrical resistivity data under applied pressure 
and magnetic field (inset of panel B). Under finite pressure, the
material shows an upturn of the electrical resistivity with decreasing
temperature, {\it i.e.}, it ceases to be metallic. We thus anticipate that
PrFe$_4$P$_{12}$ under finite pressure and PrBi under ambient pressure may
fulfill the theoretical requirements for topological two-channel Kondo
systems. Finally, UPt$_3$Au$_2$ shows 
a moderately enhanced Sommerfeld coefficient of 60\,mJ/mol/K$^2$, and an only very weakly 
temperature-dependent electrical resistivity \cite{Qui88.1} (see Fig.\,S10),
 that may point to the presence of Kondo interaction in a reduced carrier concentration
 setting.
 
 \subsection*{M. Space group symmetry constraints in the candidate materials}
	As discussed in the main text, PrBi \cite{PrBi_dft,PrBi_exp} and PrFe$_4$P$_{12}$ 	
\cite{PrFe4P12_exp,PrFe4P12_theory} have space groups $\#225$ and	
$\#204$, respectively. Whereas	
UPt$_3$Au$_2$ \cite{Qui88.1}, which has a broken inversion symmetry, 	
has space group $\#216$.	
These space groups allow for Dirac or Weyl points.	
For example, in PrBi, if two bands with the crossing point 
on the $\Gamma$-$X$ line have different eigenvalues under 	
the 
$4^+_{100}$ rotational symmetry, 
it would be a symmetry-protected Dirac node; this symmetry-based argument is manifested in the 	
DFT calculations (albeit without treating correlation effects), in which	
a band crossing occurs on the $\Gamma$-$X$ line ~\cite{PrBi_dft}. 	
Similarly, the $3^+_{111}$ rotational symmetry could protect Dirac nodes 	
on the $\Gamma$-$H$ line in PrFe$_4$P$_{12}$. Likewise,	
the $2_{100}$ rotational symmetry could protect Weyl nodes on the $X$-$W$ line 	
of UPt$_3$Au$_2$.

 \subsection*{N. DFT calculations}
 
 The theoretical electronic-structure calculations for PrBi
 were carried out using the Vienna ab initio simulation package (VASP) 
 \cite{Kresse93.3,KRESSE199615}
 with the modified Becke-Johnson potential
 \cite{becke2006simple,Tran09.3}
 for the generalized gradient approximation of the exchange-correlation functional. 
 The pseudo-potentials in the projector augmented-wave approximation
 \cite{Kresse99.3}
included $p$ and $s$ valence electrons for Bi atoms, while for  Pr the $f$ electrons were frozen in the core. 
 The spin-orbit interaction was also considered in the calculation by the second variation method
 \cite{Hobbs00.3}.
 The self-consistent calculation of the ground state density was performed in a $11 \times 11 \times 11$ Monkhorst-Pack grid with an energy cutoff of $600$ eV.

\newpage


\section*{Supplementary Figures}

\vspace{0.5cm}

\begin{figure}[ht!]

\centering
\includegraphics[width=1\textwidth]{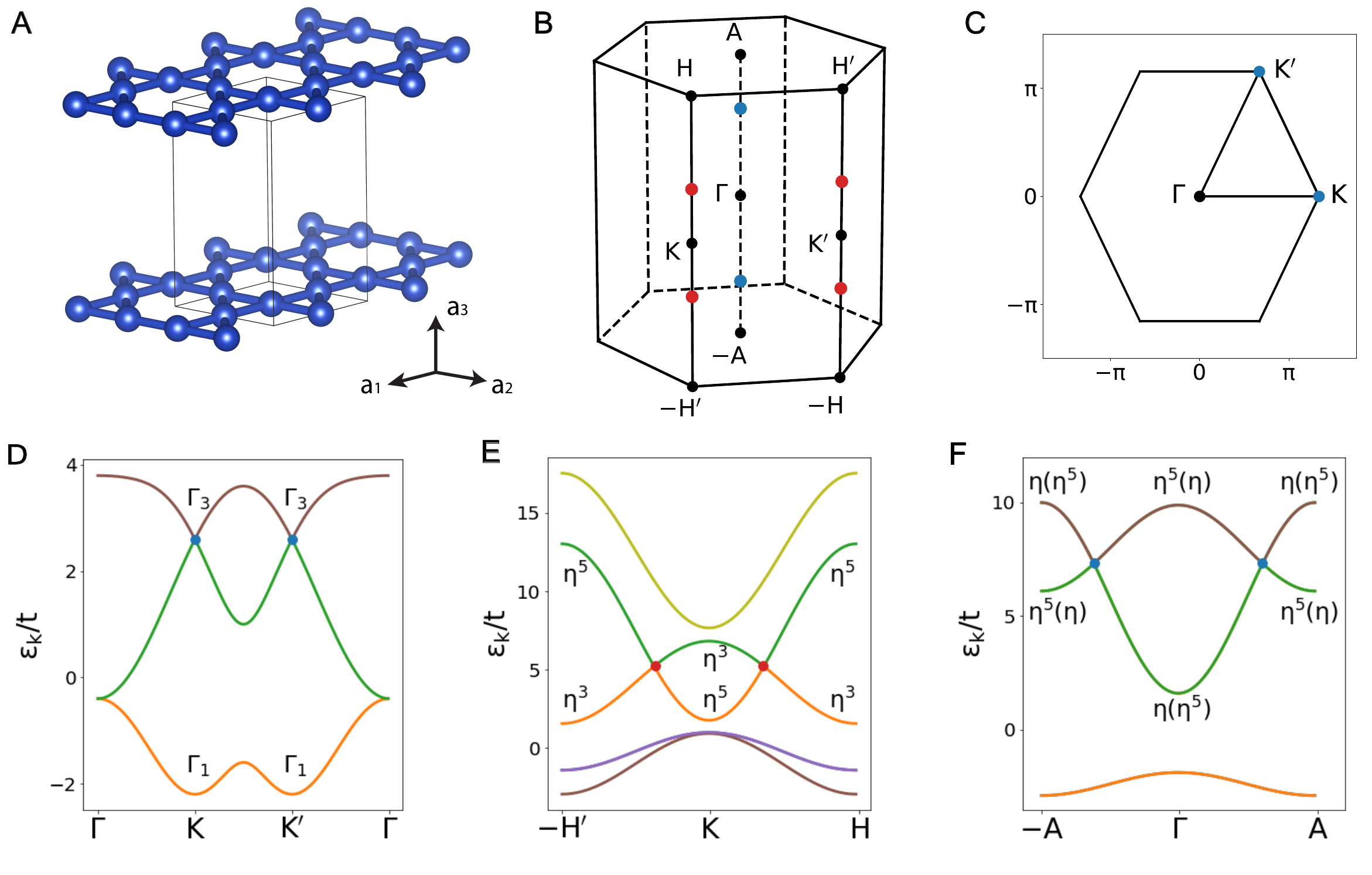}


\caption*{
\baselineskip24pt
Figure S1:  {\bf Lattice structure and the noninteracting dispersion.} ({\bf A}) The 3D kagome lattice. 
({\bf B}) The first Brillouin zone of the 3D kagome lattice, where red and blue dots denote the 
Weyl nodes and Dirac nodes, respectively. ({\bf C}) The first Brillouin zone of  the 2D kagome lattice, where the blue dots denote the 
Dirac nodes. ({\bf D}) Dispersion of the 2D kagome lattice. ({\bf E}),\,({\bf F}) Dispersion of the 3D kagome lattice along two high-symmetry lines of the Brillouin zone.
Here $0$ marks the Fermi energy of the conduction electrons in the absence of the Kondo hybridization.
The relevant bands are marked by their eigenvalues of the $C_{3z}$ symmetry, where $\eta=e^{i2\pi/6}$.
}
\label{fig:disp}
\end{figure}

\clearpage

\newpage


\begin{figure}[ht!]


    \centering
    
    \includegraphics[width=1\textwidth]{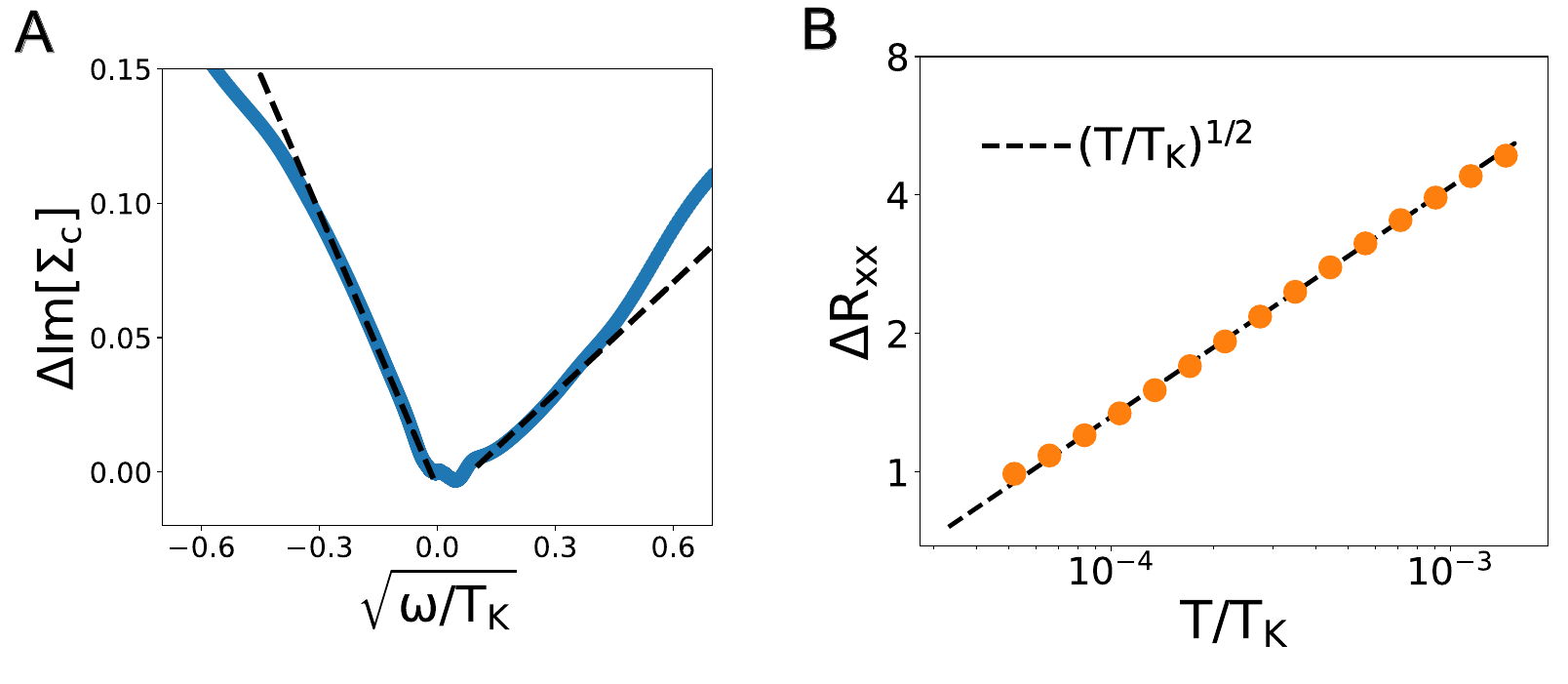}    
    \vspace{1cm}
 
   
    \caption*{
    \baselineskip24pt
Figure S2: 
    {\bf Breakdown of Landau quasiparticles}.
	({\bf A}) The self-energy ($\Sigma_c$) shows a $\sqrt{|\omega|}$-dependence 
	at
	 low temperatures (shown here for $T/T_K=10^{-3}$) in the 3D kagome lattice. 
	To isolate the frequency dependence, 
	the zero frequency contribution has been subtracted in the plot: 
	$\Delta \text{Im}[\Sigma_c](\omega) = \text{Im}[\Sigma_c](\omega)- \text{Im}[\Sigma_c](0)$.
	({\bf B}) Temperature dependence of the electrical resistivity along the $x$ direction 
	in the 3D kagome lattice. 
	Both the $\sqrt{|\omega|}$ and
	$\sqrt{T}$ dependencies characterize
	the breakdown of quasiparticles.
 }
    \label{fig:fig_sigma-c}
\end{figure} 

\clearpage

\newpage

\begin{figure}[ht!]
\vspace{-1cm}

\centering
\includegraphics[width=0.8\textwidth]{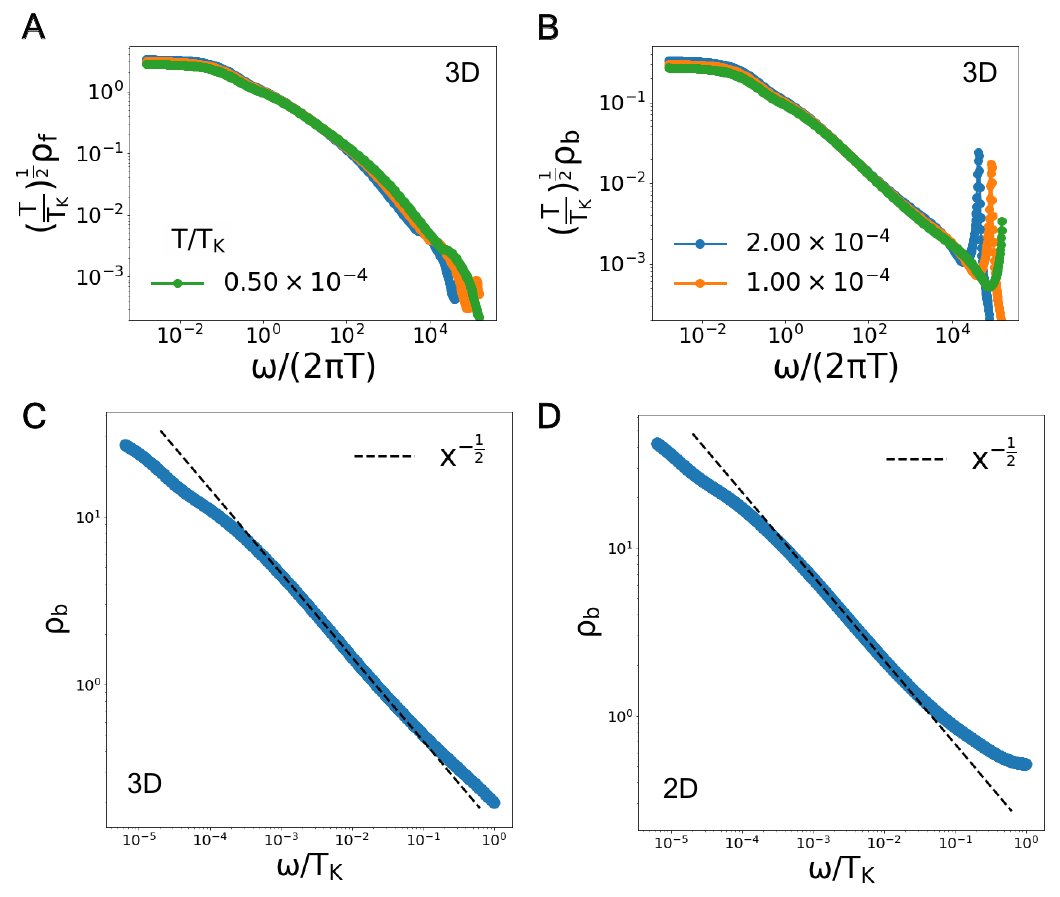}
\vspace{0.5cm}


\caption*{
\baselineskip24pt
Figure S3: {\bf Dynamical scaling and power-law behavior}.
 ({\bf A}),\,({\bf B}) spectral functions of $\rho_f$ and $\rho_b$ in the 
3D kagome lattice at various temperatures in unit of $T_K$. 
({\bf C}) and ({\bf D}) show the power-law dependence of the the pseudo-particle spectral functions 
for the models on the 3D and 2D kagome lattices, respectively. 
The spectral function $\rho_b$ is shown to depend on frequency in a power-law form,
$\omega^{-1/2}$.
The power-law form operates 
for about two decades of frequency. The deviation at high frequencies 
captures the effect of the ultraviolet energy cutoff. That at low frequencies reflects a nonzero
(albeit very low) temperature where the calculation is performed ($T=4\times 10^{-5}$ $T_K$).
}
\label{fig:slave_particle}
\end{figure}

\clearpage

\newpage

\begin{figure}[ht!]
\vspace{-1cm}

\centering
\includegraphics[width=0.8\textwidth]{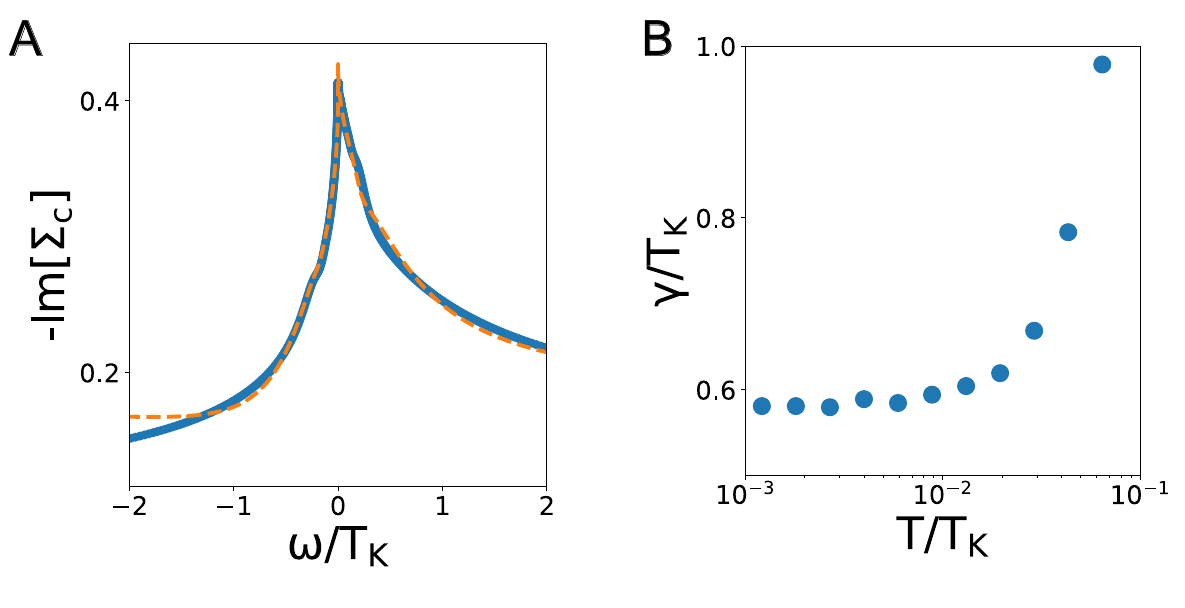}
\vspace{0.5cm}

 
\caption*{
\baselineskip24pt
Figure S4: {\bf 
Emergence of a pole in the self-energy.
}
Fitting of the self-energy 
using Eq.\,\ref{eq:sigc_MainText}.
Note that
$g(\omega)$, which describes the asymptotic behavior of the self-energy,
takes the form $a+c\sqrt{|\omega|}$.
The blue solid line shows the imaginary part of the self-energy calculated 
in the 3D kagome lattice at 
$T/T_K=10^{-3}$. The orange dash line is the fitted curve.
We observe a good consistency between the two.
({\bf B}) 
Evolution of the fitted width 
$\gamma$ as a function of temperature in the 3D kagome lattice.
It saturates to a nonzero value at temperatures low compared to $T_K$.
}
\label{fig:self_energy_fit}
\end{figure}

\clearpage

\newpage

\begin{figure}[ht!]
\vspace{-1cm}

\centering
\includegraphics[width=1\textwidth]{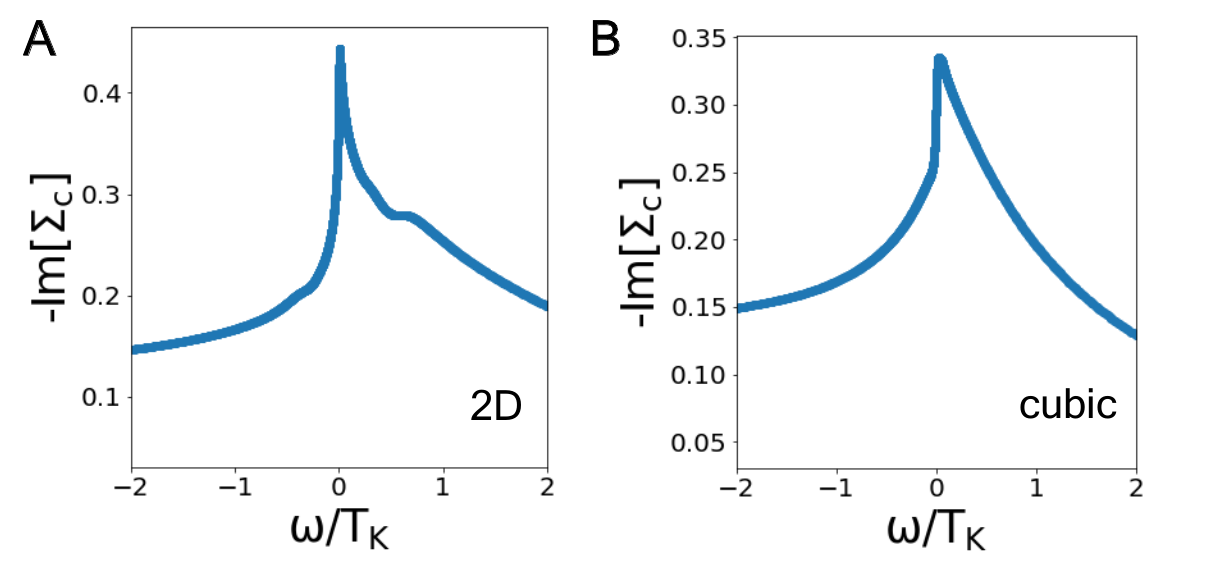}
\vspace{0.5cm}


\caption*{
\baselineskip24pt
Figure S5: {\bf 
The imaginary part of the self-energy}. 
Shown are the results for the 
2D kagome lattice 
({\bf A})
and 
cubic lattice 
({\bf B}).
The calculations are done at the temperature 
$T/T_K=10^{-3}$
 in 
 both cases.
}
\label{fig:sigc}
\end{figure}

\clearpage

\newpage

\begin{figure}[ht!]
\vspace{-1cm}

\centering
\includegraphics[width=\textwidth]{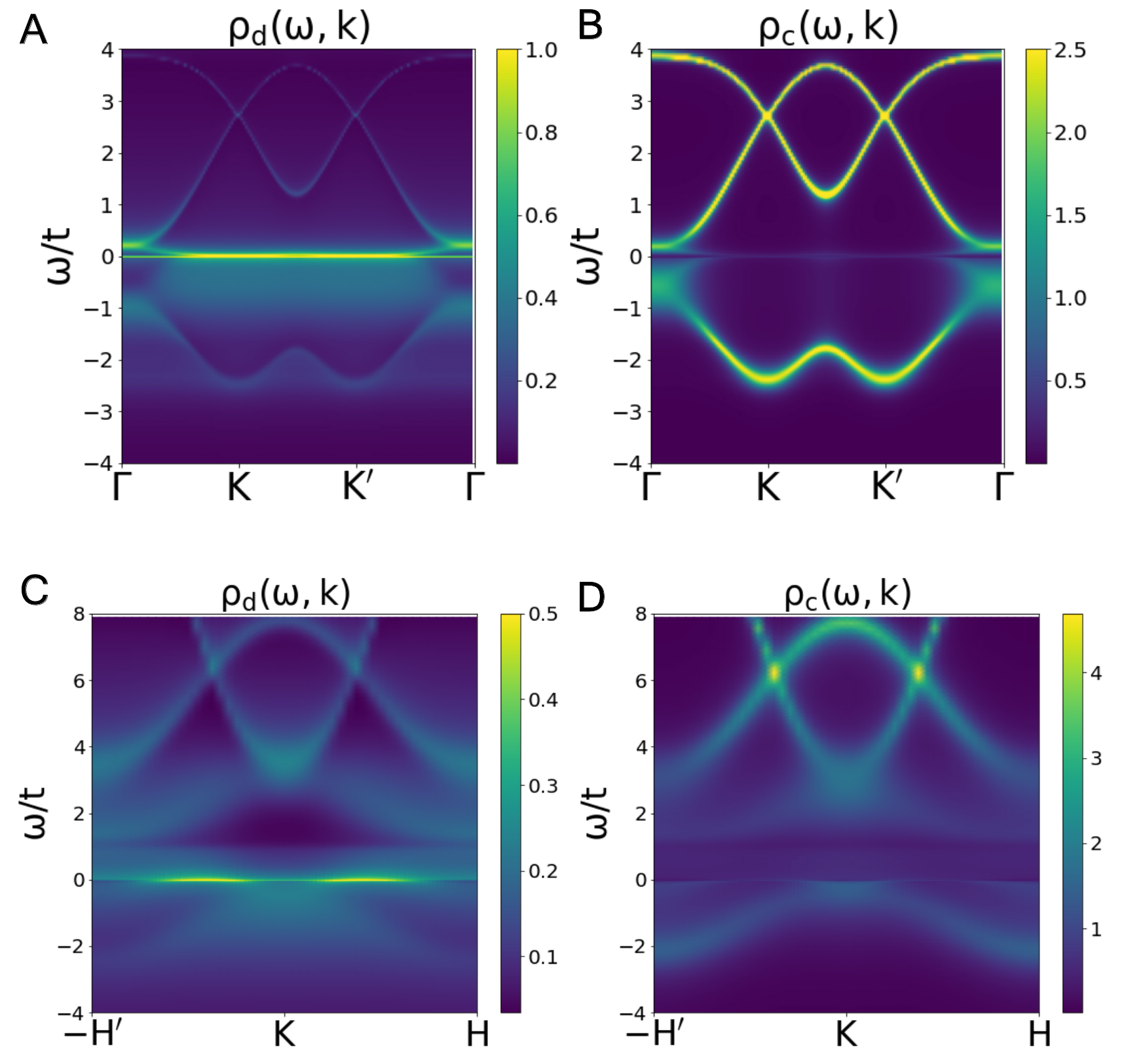}


\caption*{
\baselineskip24pt
Figure S6: {\bf Band reconstruction 
from the self-energy pole.} (\textbf{A}) and (\textbf{B}) show the spectral functions of
the  $d$ and $c$ electrons in 
the 2D kagome lattice.
(\textbf{C}) and (\textbf{D}) show the
corresponding results in 
the 3D kagome lattice. All calculations are done at $T/T_K=10^{-3}$. 
We can observe band reconstruction near the Fermi energy in both cases.}
\label{fig:spec}
\end{figure}

\clearpage

\newpage

\begin{figure}[ht!]
\vspace{-1cm}
\centering
\includegraphics[width=\textwidth]{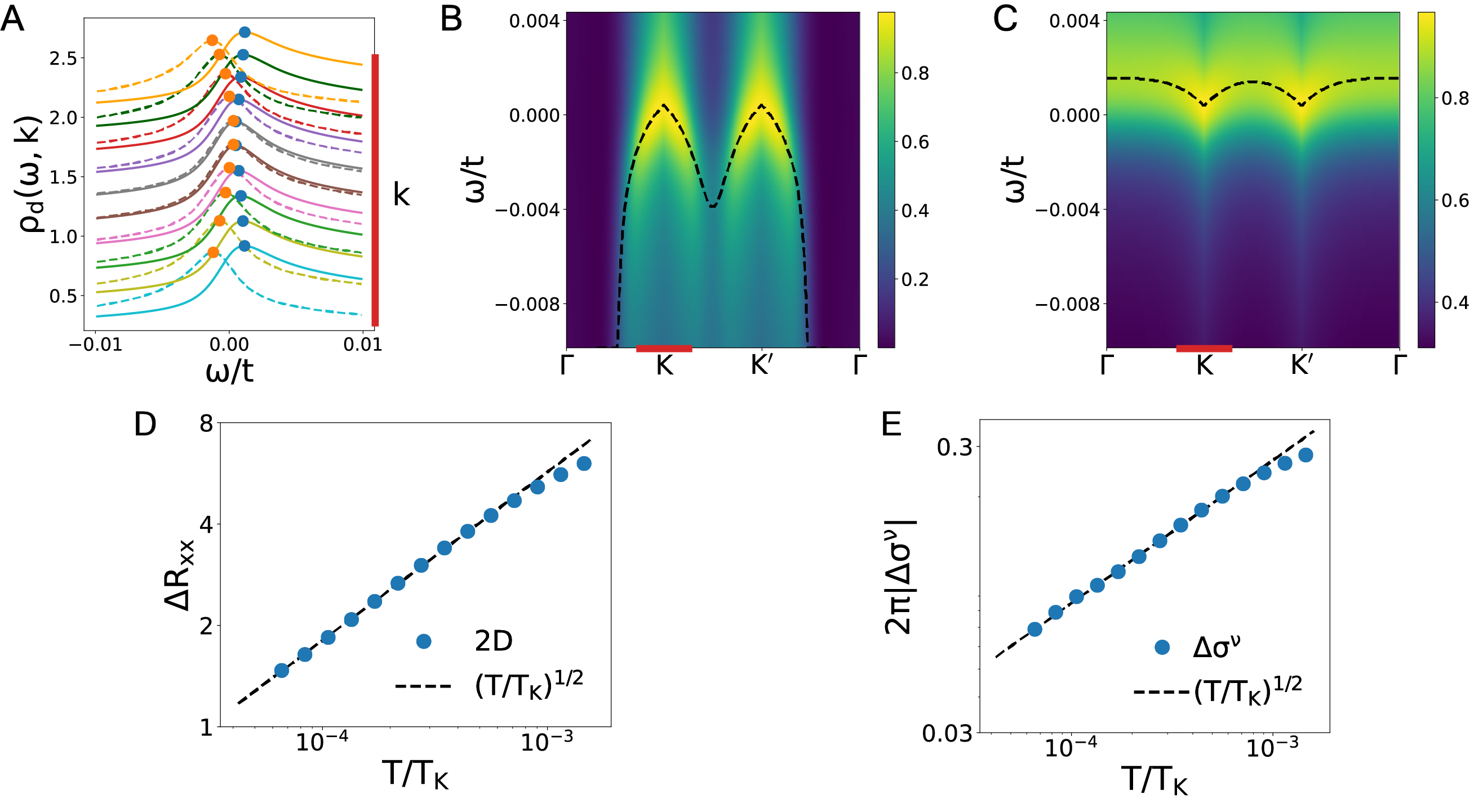}

 
\caption*{
\baselineskip24pt
Figure S7: {\bf Emergence of 
    non-Fermi liquid
    Dirac nodal excitations in the 2D model.}
    ({\bf A}) The $d$ electron spectral functions 
    ($\rho_d$)
    at various $\bm{k}$ points 
    (marked by the red bar on the right)
    along high symmetry line in the 2D kagome lattice. 
    Solid and dashed lines denote two dispersive modes from two eigenvalues of the Green's function.
    The blue and orange dots label the positions of the spectral peaks, which meet at a Dirac point.
    Each $\rho_d$
    curve
     has been shifted vertically to avoid overlapping. 
    ({\bf B}),({\bf C}) illustrate the spectral functions of the two Dirac-point-bearing branches. Here,
     the dashed line denotes the energy spectral peaks. 
     The red bar marks the cut of wave vectors in the Brillouin zone
     along which the spectral functions are shown in panel A.
    The calculations are done at the temperature
     $T/T_K=10^{-3}$ in all cases. 
Temperature dependence of the electrical resistivity along the $x$ direction ({\bf D}) 
and valley Hall conductivity ({\bf E}).
Both quantities depend on temperature in a $\sqrt{T}$ manner.
     }
\label{fig:dirac_disp}
\end{figure}

\clearpage

\newpage

\begin{figure}[ht!]
\vspace{-1cm}

\centering
\includegraphics[width=1\textwidth]{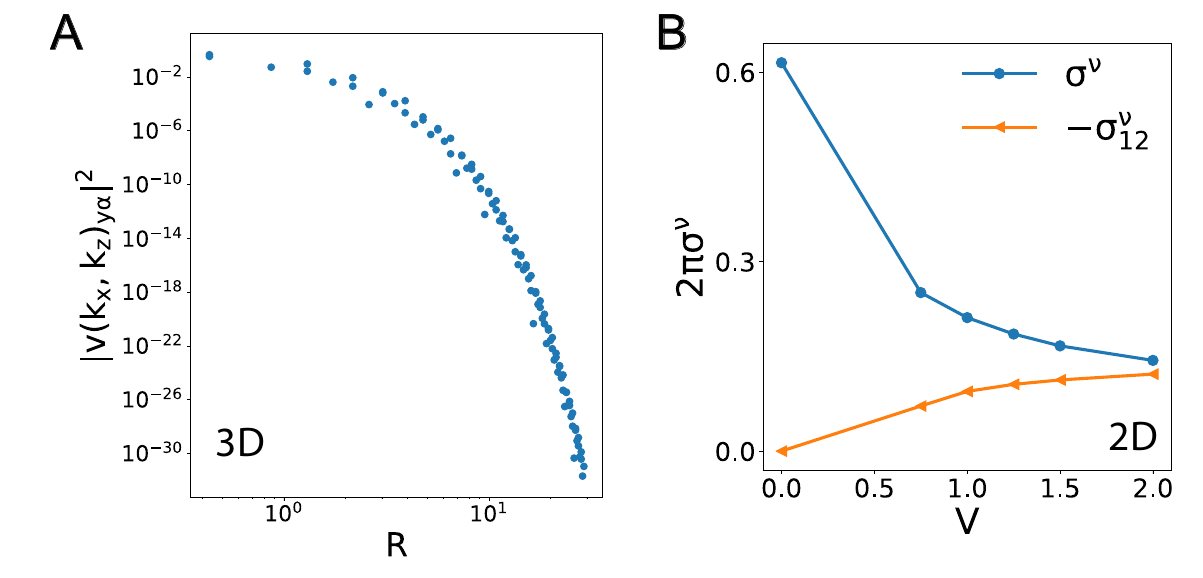}

 
 \caption*{
\baselineskip24pt
Figure S8: {\bf
Edge state and the valley Hall effect.
}
({\bf A}) Exponential decay of the edge-mode eigenvector $v(k_x=4\pi/3,k_z=0)_{\alpha y}$
in the 3D kagome model. $R$ denotes the distance to the edge.
({\bf B}) Valley Hall conductivity as a function of $V$ in the 2D kagome model 
with fixed $V/\epsilon_f=-0.5$ at $T/T_K=10^{-3}$.
}
\label{fig:topology}
\end{figure}

\clearpage

\newpage

\begin{figure}[ht!]
\vspace{-1cm}

\centering
\centerline{\includegraphics[height=0.45\textwidth]{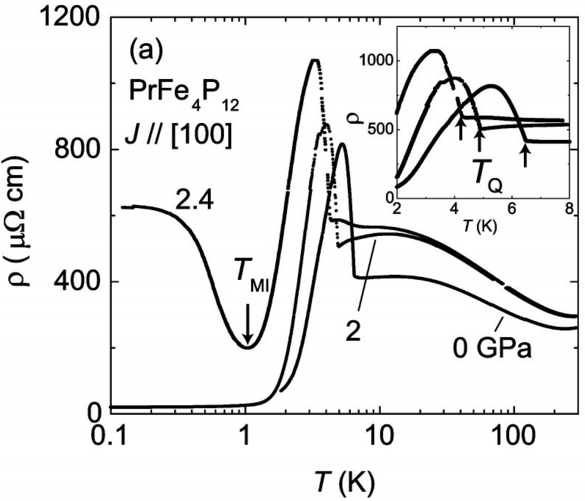}\hspace{0.4cm}\includegraphics[height=0.45\textwidth]{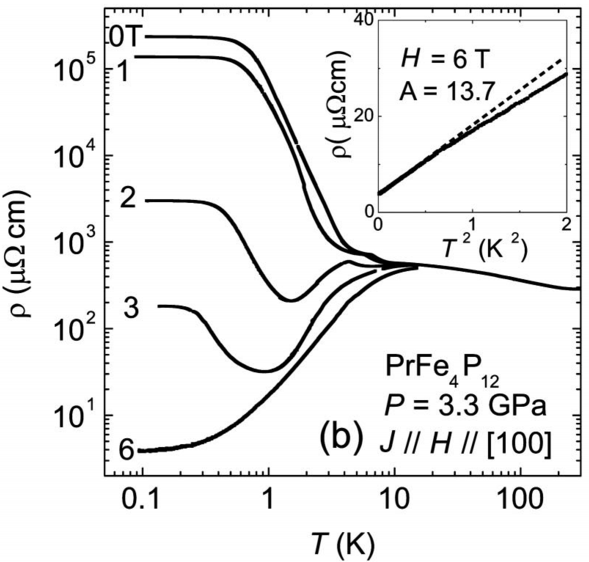}}
\vspace{-7.5cm}
\hspace{-7.3cm}{\large\bf{A \hspace{8cm} B}}
\vspace{7cm}

 
 \caption*{
\baselineskip24pt
Figure S9: {\bf Candidate two-channel
Kondo material
PrFe$_4$P$_{12}$.}  ({\bf A}) Temperature-dependent electrical resistivity at
ambient pressure and at 2\,GPa and 2.4\,GPa, revealing that the transition
temperature $T_{\rm Q}$ of the phase with antiferroquadrupolar order is
successively suppressed with increasing pressure (see inset). At lower
temperatures, a resistivity upturn is observed below $T_{\rm MI}$. ({\bf B})
Temperature-dependent electrical resistivity at 3.3\,GPa, for several fixed
magnetic fields up to 6\,T. The non-metallic state is readily suppressed by
magnetic field. At 6\,T, PrFe$_4$P$_{12}$ has metallized and exhibits heavy
fermion behavior, as evidenced by the strongly enhanced $T^2$ prefactor
$A=13.7\,\mu\Omega$cm/K$^2$ which, assuming the validity of the Kadowaki-Woods
relation, corresponds to a Sommerfeld coefficient of about 1\,J/mol/K$^2$, in
good agreement with specific heat experiments \cite{Sug02.1}. From
Ref.\,\citenum{PrFe4P12_exp}.
}
\label{fig:materials}

\end{figure}

\clearpage

\newpage

\begin{figure}[ht!]
\vspace{-1cm}

\centering
\centerline{\includegraphics[height=0.45\textwidth]{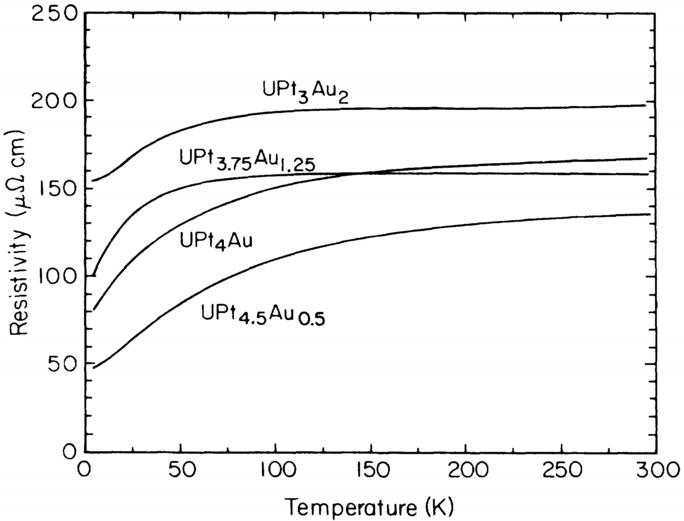}}

 
 \caption*{
\baselineskip24pt
Figure S10: {\bf Candidate two-channel 
Kondo material
UPt$_3$Au$_2$.}  ({\bf A}) Temperature-dependent electrical resistivity for  several UPt$_{5-x}$Au$_x$ samples. The weak temperature dependence of UPt$_3$Au$_2$, together with the large resistivity values at low temperatures, suggests that it is not metallic. From Ref.\,\citenum{Qui88.1}.
}
\label{fig:materials}

\end{figure}

\clearpage

\newpage

\begin{figure}[ht!]
\vspace{-1cm}
\centering
\centerline{\includegraphics[height=0.70\textwidth]{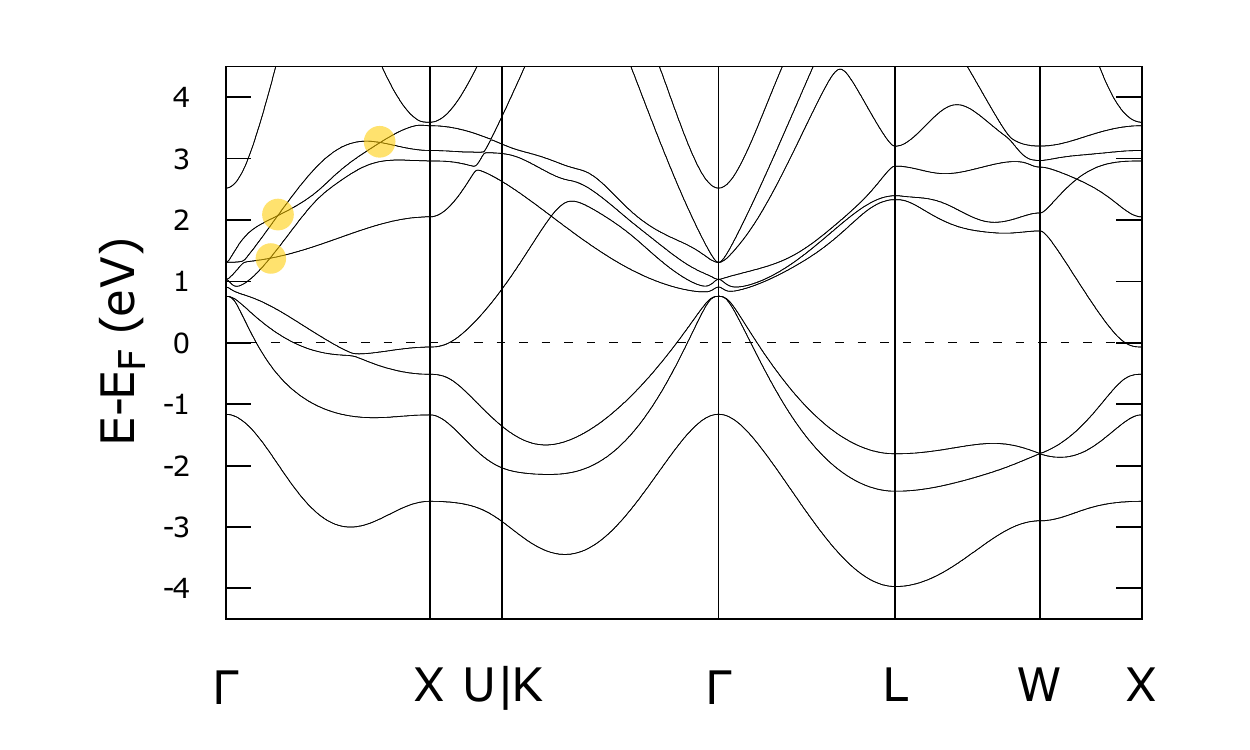}}

 
 \caption*{
\baselineskip24pt
Figure S11: {\bf DFT results in PrBi.} 
The bandstructure of the $spd$ conduction electrons of PrBi, from $f$-core 
DFT calculations. 
The three yellow dots mark the Dirac points 
The Dirac points in the non-$f$ conduction electron states
appear far away from the Fermi energy,
and support the lattice symmetry argument given in this SM, Sec.\,M.
}
\label{fig:dft_prbi}

\end{figure}

\end{document}